\documentclass[aps,pra,twocolumn,showpacs,superscriptaddress,10p,longbibliography]{revtex4-2}
\usepackage{amssymb}
\usepackage{graphicx}
\usepackage[sort&compress]{natbib}
\usepackage[colorlinks = true,linkcolor = red,citecolor = blue,urlcolor=blue]{hyperref}
\usepackage{physics}
\usepackage{scalerel}
\usepackage[normalem]{ulem}
\usepackage[dvipsnames]{xcolor}
\usepackage{bm}
\usepackage{amsmath}
\usepackage{subfigure}
\usepackage{amsfonts}
\usepackage{mathtools}
\usepackage{dsfont}
\usepackage{float}
\usepackage{lmodern}
\usepackage[many]{tcolorbox}
\usepackage{empheq}
\usepackage{soul}
\usepackage[T1]{fontenc}

\usepackage[capitalise]{cleveref}

\newcommand{\av}[1]    {\langle #1 \rangle}
\newcommand{\modsq}[1]    {\left| #1 \right|^2}

\usepackage{tikz}
\definecolor{lime}{HTML}{A6CE39}
\DeclareRobustCommand{\orcidicon}{
    \begin{tikzpicture}
    \draw[lime, fill=lime] (0,0) 
    circle [radius=0.13] 
    node[white] {{\fontfamily{qag}\selectfont\tiny ID}};
    \draw[white, fill=white] (-0.0625,0.095) 
    circle [radius=0.007];
    \end{tikzpicture}
    \hspace{-2mm}}
    
\newcommand{\orcidPS}{\href{0000-0001-9219-7274}{\orcidicon}}

\newcommand{\orcidJK}{\href{https://orcid.org/0000-0003-0998-9460}{\orcidicon}}

\newcommand{\orcidMP}{\href{https://orcid.org/0000-0002-0835-1644}{\orcidicon}}

\newcommand{\orcidTHY}{\href{https://orcid.org/0000-0001-8982-9277}{\orcidicon}}

\begin{document}

\title{Non-stabilizerness in quantum-enhanced metrological protocols
}

\author{Tanausú Hernández-Yanes
\orcidTHY}
\affiliation{Instytut Fizyki Teoretycznej, Wydzia\l{} Fizyki, Astronomii i Informatyki Stosowanej, Uniwersytet Jagiello\'nski, \L{}ojasiewicza 11, PL-30-348 Krak\'ow, Poland}

\author{Piotr Sierant\orcidPS}
\affiliation{Barcelona Supercomputing Center, Barcelona 08034, Spain}

\author{Jakub Zakrzewski\orcidJK}
\affiliation{Instytut Fizyki Teoretycznej, Wydzia\l{} Fizyki, Astronomii i Informatyki Stosowanej, Uniwersytet Jagiello\'nski, \L{}ojasiewicza 11, PL-30-348 Krak\'ow, Poland} 
\affiliation{Mark Kac Complex Systems Research Center, Jagiellonian University in Krak\'ow, PL-30-348 Krak\'ow, Poland}

\author{Marcin Płodzień\orcidMP}
\affiliation{Qilimanjaro Quantum Tech, Carrer de Veneçuela 74, 08019 Barcelona, Spain} 
\affiliation{ICFO-Institut de Ciencies Fotoniques, The Barcelona Institute of Science and Technology, 08860 Castelldefels (Barcelona), Spain}

\date{\today}

\begin{abstract}
{Non-stabilizerness (colloquially "magic") characterizes genuinely quantum (beyond-Clifford) operations necessary for preparation of quantum states, and can be measured by stabilizer R\'enyi entropy (SRE). For permutationally symmetric states, we show that the SRE depends, for sufficiently large systems, only on a constant number of expectation values of collective spin operators. This compact description is leveraged for analysis of spin-squeezing protocols, which inherently generate non-stabilizerness. 
{Under one-axis twisting (OAT), the generation of optimal squeezing is accompanied by a logarithmic divergence of SRE with system system size. Continued time evolution under OAT produces metrologically useful ‘kitten’ states—superpositions of rotated GHZ states—that feature many-body Bell correlations but exhibit a smaller, system-size-independent SRE that decreases with increasing Bell-correlation strength. Our results reveal connections between non-stabilizerness, multipartite correlations, and quantum metrology, and provide a practical route to quantify non-stabilizerness in experiments for precision sensing.}
}
\end{abstract}

\maketitle

\section{Introduction}
Quantum technologies \cite{Dowling2003SecondQuantumRevolution} draw their power from physical resources~\cite{Chitambar19,Salazar2021,Salazar2024,Kuroiwa2024} absent in classical {systems}.  Quantum entanglement \cite{Horodecki2009, Horodecki2024} and Bell-type correlations \cite{Brunner2014} enable unconditional security in communication
\cite{BennettBrassard1984,Ekert1991,LoChau1999,Gisin2002,Scarani2009,Xu2020}, exponential compression in simulation~\cite{Georgescu14}, and measurement sensitivities beyond the shot-noise limit in metrology~\cite{Giovannetti2006QuantumMetrology,Pezze2018RMP,Degen2017QuantumSensing,Agarwal2025}, while coherence \cite{Streltsov2017} provides the temporal substrate on which these correlations are created, manipulated, and read out.  Together, these ingredients have fueled rapid progress across quantum science and technology.  
However, entanglement and coherence alone are not sufficient to unlock the computational power of quantum systems. Stabilizer states~\cite{Gottesman98},
despite containing large amounts of entanglement and coherence, can still be efficiently represented on classical computers~\cite{Aaronson04}. This motivated the introduction of the magic state resources, or \emph{non-stabilizerness}~\cite{veitchNegativeQuasiprobabilityResource2012, veitchResourceTheoryStabilizer2014, Bravyi16, Howard17, Wang19, Heimendahl21, Magni25, Magni25complexity}, which quantifies the distance of a quantum state from the set of classically simulable stabilizer states. A quantum device can achieve a computational quantum advantage~\cite{Preskill12, Daley22} only if its state contains a sufficient amount of non-stabilizerness.

\begin{figure}[t!]
    \centering
    \includegraphics[width=\linewidth]{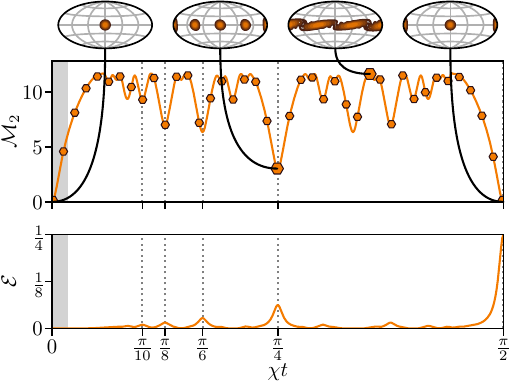}
        \vspace{-0.7cm}
    \caption{
    Dynamics of SRE, $\mathcal{M}_2$ (upper panel), and many-body Bell correlator, $\mathcal{E}$ (lower panel), during OAT dynamics in the time scale $t \le \pi/2$ for $N=100$.  
    Markers indicate exact numerical SRE calculations, while solid lines show the value of $\mathcal{M}_2$ calculated with our compact formula~\cref{eq:SRE_approx}. 
    Insets show the corresponding Husimi functions at marked times. 
    Shaded areas correspond to the spin-squeezing time scale while dashed lines correspond to times at which kitten states are generated.
    }
    \label{fig:kittens_t}
\end{figure}

{
Stabilizer Rényi entropy (SRE)~\cite{leoneStabilizerRenyiEntropy2022} possesses the key features of a non-stabilizerness measure~\cite{Leone24}, can be efficiently evaluated numerically~\cite{Haug23mps, Lami23, Tarabunga23, Tarabunga24} and measured experimentally~\cite{Oliviero22, Haug23, Haug24,Ahmadi2024}. The introduction of SRE has enabled systematic investigations of non-stabilizerness across distinct quantum phases and at quantum phase transitions~\cite{White21, Odavic23, Tarabunga24rk, Falcao25, Jasser2025, Bera2025SYK}, in variational quantum states \cite{Spriggs2025,Sinibaldi2025}, as well as in out-of-equilibrium settings~\cite{Rattacaso23, Turkeshi25spectrum, Odavic25, Haug25, Kos24, Santra25, Passarelli25boundary, Hou2025highway, Turkeshi23flatness}, including random quantum circuits~\cite{Turkeshi24magic}, ergodic~\cite{Tirrito24anti}, and non-ergodic~\cite{Smith25, Falcao25mbl} many-body dynamics. While these efforts have shed light on non-stabilizerness in various model settings, its role in many-body systems that form the backbone of quantum sensors remains largely unexplored.}

Among these, the one-axis twisting (OAT) protocol~\cite{Wineland1992, Kitagawa1993} is of particular interest.
OAT can be realized with a variety of ultra-cold systems via atom-atom collisions~\cite{Riedel2010,Gross2010,Hamley2012,Qu2020} and atom-light interactions~\cite{Leroux2010, Maussang2010}. 
OAT simulation with ultra-cold atoms in optical lattices receives particular interest in recent theoretical proposals~\cite{Kajtoch2018, He2019, Plodzien2020, Mamaev2021, Comparin2022, HernandezYanes2022, Dziurawiec2023, HernandezYanes2024, HernandezYanes2025} and experimental implementations~\cite{Eckner2023,Miller2024}.
An initial product state of spins polarized along a properly chosen direction is squeezed through OAT to obtain scalable many-body entangled and Bell correlated states~\cite{Sorensen2001,Sorensen2001Nature,Wang2003PRA, Korbicz2005, Hyllus2012PRA, Vitagliano2014, Tura2014, Schmied2016, Aloy2019, Baccari2019, Tura2019, MullerRigat2021, Plodzien2022, Plodzien2024PRR}. 
In particular, OAT generates spin-squeezed states with phase sensitivities approaching the Heisenberg limit~\cite{Wineland1994, Sinatra2011, Ma2011, Pezz2018, Sinatra2022}.

In this work, we investigate the birth and growth of non-stabilizerness during OAT dynamics for $N$ (even) spin-$1/2$ particles, aiming to characterize the non-stabilizerness of analog simulation protocols.
We consider the SRE to quantify the magic state resources
and assess when and to what extent OAT generates non-stabilizerness.
We derive a closed-form expression for SRE within the permutation-symmetric sector that expresses the SRE in terms of six projections of the analyzed state. This expression, valid for sufficiently large system size $N$, enables us to understand the non-stabilizerness of the states generated in OAT protocol 
and relate it with spin squeezing and many-body Bell correlations of the generated states. 

\section{Stabilizer Rényi Entropy in the permutation-symmetric sector}
{The} SRE
of a $N$--qubit pure state \(\ket{\psi}\) is defined~\cite{leoneStabilizerRenyiEntropy2022} as 
\begin{equation}
    \label{eq:SRE_full}
    \mathcal{M}_q(\ket{\psi})=\frac{1}{1-q}\log_2\!\left(\frac{1}{2^N}\sum_{\hat{P}\in\mathcal{P}_N}\!\!\langle 
    \psi|\hat{P}| \psi \rangle^{2q}\right),
 \end{equation}
where $q>0$ is the R\'{e}nyi index,  $\mathcal{P}_N=\{ \hat{P}\}$ is the group of $4^N$ Pauli strings, i.e. operators $\hat{P} = \hat\sigma_1\otimes\cdots\otimes\hat\sigma_N$ that are tensor products of identity and Pauli $X, Y, Z$ operators, $\hat{\sigma_i}= \hat{\mathbb{I}}_i, \hat{X}_i, \hat{Y}_i, \hat{Z}_i$, modulo $\{ \pm1, \pm i\}$ phase. 
The Clifford group $C_N=\{U_C\}$ is a subgroup of the unitary group $U(2^N)$ that maps Pauli string to Pauli string, and stabilizer states, $\ket{\psi_{\mathrm{stab}}} =U_C \ket{0}^{\otimes N}$ are generated by action of $C_N$ on computational basis state $\ket{0}^{\otimes N}$.
As a measure of non-stabilizerness, the SRE is faithful, i.e. $\mathcal{M}_q(\ket{\psi})=0$ if and only if $\ket{\psi}$ is a stabilizer state; Clifford invariant, $\mathcal{M}_q(U_C \ket{\psi})=\mathcal{M}_q(U_C \ket{\psi})$; and additive on product states, $\mathcal{M}_q(\ket{\psi}\otimes \ket{\phi}) = \mathcal{M}_q(\ket{\psi})+\mathcal{M}_q(\ket{\phi})$. Moreover, the SRE is a pure state non-stabilizerness monotone~\cite{Leone24}, meaning that the SRE is non-increasing under stabilizer protocols such as partial trace, computational basis measurements, or composition with $\ket{0}$, provided that the state of the system remains pure.

Projecting the Pauli strings onto the permutation-symmetric  sector~\cite{MullerRigat2025} as done by Passarelli et al. \cite{Passarelli2024} {(see also~\cite{Passarelli25})} 
{reduces the computational complexity of SRE evaluation from exponential to polynomial in  $N$. Still, {computing} the SRE  requires evaluation of the expectation values of $O(N^3)$ distinct Pauli-string representatives. }
We describe  {the permutation-symmetric} sector in a different basis to show that the SRE in the limit $N\to\infty$  {is determined by} projections {of the state} on {to} a few relevant stabilizer states.

While the permutation-symmetric sector is spanned by the angular momentum eigenstates with $J=N/2$ $\ket{J = N/2, m}; \forall m \in [-J,J]$, also referred to as Dicke states, states from this {sub}space can also be expressed using 
 the overcomplete basis of spin coherent states 
 $
	\ket{\theta, \phi} 
	= \bigotimes_{j=1}^{2J} \left( \cos({\theta}/{2})\ket{0} + e^{-i\phi}\sin({\theta}/{2})\ket{1} \right) 
    = \sum_{m} \sqrt{2J \choose J+m} \cos^{J-m} (\theta / 2) (\sin(\theta / 2) e^{-i\phi})^{J+m}\ket{J, m}
    $. 
From this definition and the completeness of the angular momentum eigenstates we can resolve the identity operator in terms of the spin coherent states as \cite{Klauder1985, Robert2021, Kam2023}
$  \hat{\mathbb{I}}_{J} 
    =\frac{2J+1}{4\pi}\int d\vb*{\Omega} \ket{\theta,\phi}\bra{\theta,\phi}$,
where $d\vb*{\Omega}$ is the differential solid angle in spherical coordinates.

 {Since the SRE, \cref{eq:SRE_full} is determined by the expectation values of Pauli strings $\hat{P}$, and we are interested in the permutation-symmetric sector, we}   project  $\hat{P}$  {onto the} spin coherent states basis as

\begin{multline}
    \label{eq:pauli_arbitrary}
        \hat{\mathbb{I}}_{J}\hat{P}\hat{\mathbb{I}}_{J} 
		= \left( \frac{2J+1}{4\pi} \right)^2 \int \int d\vb*{\Omega} d\vb*{\Omega'} 
		 \ket{\theta, \phi} \bra{\theta', \phi'} 
	\\
	\cdot 
	\bra{\theta, \phi} \hat{P} \ket{\theta', \phi'}
.\end{multline}
{The main contributions to the operator in \cref{eq:pauli_arbitrary} come from the
matrix elements $\bra{\theta, \phi} \hat{P} \ket{\theta', \phi'}$ of $\hat{P}$ of the form $\bra{\pm \sigma} \hat{P} \ket{\pm \sigma}$, $\bra{\pm \sigma} \hat{P} \ket{\mp \sigma}$, where $\ket{\pm \sigma}$ are spin coherent stabilizer states. The spin coherent stabilizer states, with $\sigma=X,Y,Z$, correspond to $\ket{ \pm X} \equiv \ket{\pi /2, \pi/2 \mp \pi/2}$, $\ket{\pm Y} \equiv \ket{\pi /2,  \pm \pi / 2} $,  $\ket{\pm Z} \equiv \ket{\pi /2 \pm \pi / 2, 0}$. Contributions from any other matrix element $\bra{\theta, \phi} \hat{P} \ket{\theta', \phi'}$ decay exponentially as $\alpha^{N_X} \beta^{N_Y} \gamma^{N_Z} \kappa^{N_I}$ where $N_X, N_Y, N_Z, N_I$ are non-negative integers fulfilling the constraint $N_X+N_Y+N_Z+N_I = N$ and 
$x_1, x_2, x_3 < 1$, $x_4 \le 1$ for $x_i \in\{|\alpha|,|\beta|,|\gamma|,|\kappa|\}$, see \cref{sup:pauli_spectra,sup:pauli_solutions_1100}. 
}
 
We apply these observations in \cref{eq:SRE_full} to obtain our main result: the approximation for SRE in the {$N\to\infty$} limit for permutation-invariant states:
\begin{equation}
	\lim_{N \to \infty} \mathcal{M}_q(\ket{\psi}) 
	\simeq \frac{1}{1-q}\log_2\left( \frac{1}{2}
    \sum_{\sigma, n, m}
    \left(c_{n,m}^{(\sigma)}\right)^{2q} \right),
    \label{eq:SRE_approx}
\end{equation}
where ${\sigma\in\{X,Y,Z\}}$, ${n\in\{1,2\}}$, ${m\in\{1,2\}},$
with coefficients $c^{(\sigma)}_{1,m} = |\bra{\sigma}\ket{\psi}|^2 + (-1)^m |\bra{-\sigma}\ket{\psi}|^2$, $	c^{(\sigma)}_{2,m} = \sqrt{(-1)^m}(\bra{\psi}\ket{-\sigma}\bra{\sigma}\ket{\psi} +(-1)^m \bra{\psi}\ket{\sigma}\bra{-\sigma}\ket{\psi})$,
which \emph{only} depend on the six single-axis projections \(\bra{\pm X}\ket{\psi}, \bra{\pm Y}\ket{\psi}, \bra{\pm Z}\ket{\psi}\). \Cref{eq:SRE_approx} thus provides a lightweight, comprehensive, and experimentally accessible estimator for \(\mathcal{M}_q\) in the large \(N\) symmetric subspace. For derivation details, see~\cref{sup:large_N}.

\begin{figure}[t!]
    \centering    \includegraphics[width=\linewidth]{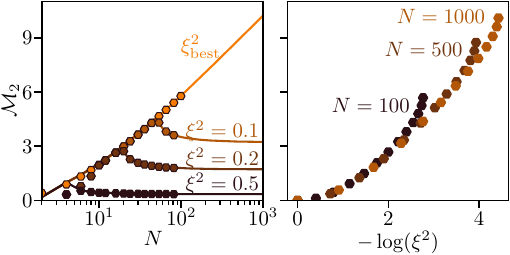}
    \caption{    
    Left panel shows scaling of $\mathcal{M}_2$ with $N$ for spin-squeezed states at fixed $\xi^2$. 
    Markers indicate exact SRE calculations, while solid lines show results for \cref{eq:SRE_approx}.
    The label $\xi^2_\mathrm{best}$ corresponds to the best squeezed state with $\xi^2_{\rm best} \propto N^{-2/3}$. Right panel shows scaling of $\mathcal{M}_2$ with $\xi^2$  { for $N = 100,500, 1000$}. 
    }
    \label{fig:fixed_squeezing}
\end{figure}

\Cref{eq:SRE_approx} accounts for the main contribution to the SRE in the large $N$ limit.
Further corrections can be included either through integration using softer constraints, or the addition of other matrix elements in the spin coherent basis; ordered by their magnitude.
The former strategy forsakes the advantage of summing over distinct sets of Pauli strings in exchange for a broader integration domain that accounts for missed contributions in the continuum.
The latter strategy maintains the discrete selection of relevant spin coherent states, but requires an increasing number of terms for further corrections.
We tested the latter strategy up to third order corrections with little improvement besides fringe cases involving additional collective rotations, where the qualitative behavior of the SRE is still well predicted without corrections.
Importantly, we observe that \cref{eq:SRE_approx} compares exceedingly well with exact numerical calculations of \cref{eq:SRE_full}, even for $N \sim 10$, for all the states explored throughout this work.

 \begin{figure}
    \centering
    \includegraphics[width=\linewidth]{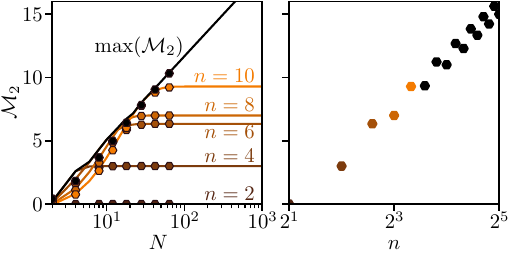}
    \vspace{-0.7cm}
    \caption{ 
    Left panel shows SRE scaling with $N$ for kitten states $\ket{\psi(\chi t = \pi / n)}$, for $n \in \{2, 4,6,8,10\}$. 
    Black line and markers show the numerically obtained maximal SRE for $\ket{\psi(\chi t \le \pi / 2)}$.
    Right panels shows SRE of particular kitten-states $n$ for $N=1000$ spins.
    }
    \label{fig:kittens}
\end{figure}

\section{Results and discussion}

We define our OAT generated quantum states  through the unitary operator $\hat{U}(t) = \exp\{-i \chi t \hat{Z}^2/4\}$, with $\hat{Z}= \sum_j \hat{Z}_j$, acting on the initial state $\ket{X}$, i.e.
\begin{equation}
    \ket{\psi(t)} 
                 = \frac{1}{2^J}\sum_{k=0}^{2J} \sqrt{{2J \choose k}} e^{-i \chi t (k-J)^2} \ket{J, k-J}.
    \label{eq:psiTEVOL}
\end{equation}

The OAT protocol yields a hierarchy of metrologically enhanced, many-body–correlated states~\cite{Plodzien2022}, each with distinct robustness. 
\emph{Spin-squeezed} states appear for $t\!\lesssim\! t_{\rm best}\!\approx\!\chi^{-1}N^{-2/3}$ and give gain tolerant to finite-resolution detection and to moderate loss/dephasing~\cite{Pezz2018}. 
At later times, \emph{oversqueezed} non-Gaussian states arise with greater entanglement depth; they resist detection noise but are sensitive to dephasing~\cite{Davis2016,Nolan2017,Baamara2021}. 
For $\chi t\!\simeq\!\pi/n$ (integer, even $n$), 
the state is a macroscopic superposition of maximally separated $n$ coherent states, i.e. a multi-head cat (\emph{kitten}) state, being relatively loss tolerant and enabling sensitivities from sub-SQL to near Heisenberg \cite{Tatsuta2019} -yet fragile to phase noise~\cite{Ferrini2008,Spehner2014,Huang2015,Chalopin2018}. 
At $\chi t\!\simeq\!\pi/2$, a GHZ state with Heisenberg scaling is produced, but it is highly vulnerable to perturbations.

To quantify the relation between SRE and the extent of many-body Bell correlations
generated in the OAT process, we use a quantum $N$-body correlator $Q$ which witnesses Bell correlation when the following inequality is violated ~\cite{ CavalcantiPRL2007,CavalcantiPRA2011,HePRA2011,Niezgoda2020,Niezgoda2021,Chwedeczuk2022,Plodzien2022,Plodzien2024PRA,Hamza2024,Plodzien2025ROPP,Plodzien2025PRA}:
\begin{equation}
    Q = \log_2(2^N {\mathcal E})<0,\quad
{\mathcal E}= \big|\frac{1}{N!}\langle\psi(t)|\hat{J}_+^N|\psi(t)\rangle\big|^2,
\end{equation}
where $\hat{J}_+ \equiv\frac{1}{2}(\hat Y +i\hat Z)$ is a rising operator in the $x$ direction. The value of ${\mathcal E}$ (equivalently $Q$) captures information on the structure of the many-body quantum state. For example, for all separable states we have $Q\le0$. On the other hand, when the system is in non-2-separable state, being a pure GHZ state, the correlator is maximized, $\mathcal{E} = \frac{1}{4}$, $Q = N-2$. 
For states being a product of a single qubit and entangled
state of the remaining $N-1$ qubits  the
correlator is maximal when the $N-1$ qubits form a GHZ state;
hence 
$
  Q \le N - 2,
$
and its violation indicates genuine {multipartite} entanglement, i.e., the state
is non-2-separable. 
For details, see \cite{Chwedeczuk2022,Plodzien2025ROPP,Plodzien2025PRA}.

\subsection{Magic state resources of spin-squeezed states}
At short time scales, $t\lesssim t_{\rm best}$, 
the evolved state is   a spin squeezed state.
Spin squeezed states {are characterized by a}
reduced variance along a given spin component with respect to the standard quantum limit, 
which enhances measurement sensitivity along this direction. 
{
Their metrological usefulness can be  {quantified} by the spin squeezing parameter $\xi^2$ derived by Wineland et al. \cite{Wineland1994},  {defined as} 
$\xi^2 = N \min_{\vb*{n}} (\Delta \hat{\sigma}_{\perp, \vb*{n}}^2 ) / \ev*{\hat{X}}^2$, where 
$\hat{\sigma}_{\perp, \vb*{n}} = \vb*{n} \cdot (\hat{Y}, \hat{Z})$ and $\vb*{n}$ is an arbitrary normalized vector.
}   
When $\xi^2$ drops below unity, it signals the presence of entanglement between the spins as well as a metrological gain relative to the standard quantum limit. For OAT, the optimal squeezing is obtained at $\chi t_{\rm best} \sim N^{-2/3}$, with minimal spin squeezing parameter value $\xi^2_{\rm best} \sim N^{-2/3}.$

Starting from the initial stabilizer state $\ket{X}$ with ${\cal M}_q = 0$, the interactions generate spin-squeezing until $t_\mathrm{best}$, producing a maximally squeezed state, with a minimal value of $\xi^2$. On this time scale, $\ket{\psi(t)}$ 
{provides a sub-binomial probability distribution with respect to the Dicke states quantized along the squeezing direction \cite{takahashiQuantumNondemolitionMeasurement1999}}, with monotonically increasing SRE; 
shown in the shaded gray area of the upper panel in \cref{fig:kittens_t}.
The solid line represents the SRE obtained using \cref{eq:SRE_approx}, which is in full agreement with the numerical calculations of the SRE in the Dicke basis, 
see \cref{sec:compare_TACT_OAT}.
In the large $N$ limit, $\mathcal{M}_2$ saturates {to a value fixed by the squeezing parameter} $\xi^2$, see Fig.~\ref{fig:fixed_squeezing}. 
{The most strongly spin-squeezed states, with squeezing parameter $\xi^2_{\rm{best}}$}, {feature a logarithmic growth of} $\mathcal{M}_2$   with system  size {$N$}. 
For the short time OAT evolution, $t\leq t_{\rm best}$
the value of the Bell correlator can be obtained analytically \cite{Plodzien2022}, and reads
$
{\mathcal E} \approx \frac{4}{(1+\kappa^2)^2} e^{-\frac{\pi^2}{8}\frac{N}{1+\kappa^2}}, \quad \kappa = \chi t\frac{N}{2}$.
We conclude that the SRE and many-body Bell correlations for the best squeezed states, at $t=t_{\rm best}$, 
$Q  \approx N -  aN^{1/3}$ (where $a = \frac{\pi^2}{2} \log_2 e$)
grows sub-linearly with $N$.

\subsection{Magic state resources of kitten states}
{For $t>$} $t_{\mathrm{best}}$,
the {time-}evolved state {$\ket{\psi(t)}$}
crosses into the \emph{oversqueezed} regime: the optimal transverse variance reverses and grows (so $\xi^{2}$  increases), and continued twisting shears the wavepacket away from Gaussianity. 
At  time instances 
$\chi t_{n}=\pi/n$ for positive even integer $n$  the state is macroscopic superposition of maximally separated  $n$ coherent states. 
{A tedious but direct inspection of Eq.~\eqref{eq:psiTEVOL} for $t= t_{n}$ shows that} the $n$-head kitten state ($n$ even) can be expressed as a superposition of rotated GHZ states, i.e.
\begin{equation}
\ket{\psi\!\left(t_{n}\right)}
= \sqrt{\frac{2}{n}}\; e^{-i\pi/4}
\sum_{s=0}^{\frac{n}{2}-1}
e^{\,i\pi s^{2}/n}\;
U_z(\phi_s)\;
\ket{\mathrm{GHZ}_x\!\big(\theta_s\big)},
\label{eq:kittenSUP}
\end{equation}
where 
$\ket{\mathrm{GHZ}_x(\theta)}
= \frac{1}{\sqrt{2}}\!\left(\ket{+X} + e^{i\theta}\ket{-X}\right)$,  and 
$U_z(\phi)=e^{-i\phi J_z}$,
$
\phi_s=\frac{2\pi s}{n}$, $s=0,1,\dots,\frac{n}{2}-1$,
$ 
\theta_s=\pi s+\frac{\pi n}{4}\quad (\mathrm{mod}\ 2\pi)$.
In particular, at $n=2$ (equivalently $\chi t=\pi/2$) systems is in a $N$-body GHZ state along $x$-direction. 
Similar states exist for odd $n$ \cite{Ferrini2008}. \cref{fig:kittens_t} shows the time evolution of SRE over half of the period of the OAT dynamics, up to $\chi t = \pi/2$.

The SRE for kitten-states does not scale with $N$, and reaches a constant value for large $N$, see Fig.~\ref{fig:kittens}.  
Each of the rotated GHZ states that span the superposition in Eq.~\eqref{eq:kittenSUP}, becomes, for $N\gg1$, a stabilizer state with a vanishing SRE,  $\mathcal{M}_q(U_z(\phi_s)
\ket{\mathrm{GHZ}_x\!\big(\theta_s\big)} )=0$. Therefore, the kitten state is a superposition $\ket{\psi\!\left(t_n\right)} = \sum_{s=0}^{n/2-1} c_i \ket{\sigma_i}$ of stabilizer states $\ket{\sigma_i}$, and its SRE, $\mathcal{M}_q(\ket{\psi\!\left(t_n\right) } )$ is bounded from above by a constant $\sum_i |c_i|$~\cite{Leone24} but independent of $N$. 

Interestingly, for even $n$-head kitten state, at large $N$, the many-body Bell correlator has simple form ${\mathcal E}\simeq n^{-2}$ \cite{Plodzien2022}. In particular, the GHZ state, $t_n =\pi/2$, maximizes the amount of many-body Bell correlations, $\mathcal{E} =1/4$, while being a stabilizer state with vanishing SRE; see \cref{fig:kittens}.  This trend persists for other kitten states: the bigger the value of SRE ($\mathcal{M}_2 \approx \ln(n)$, see the right panel of Fig.~\ref{fig:kittens}), the smaller the value of  $Q = N-2\log_2(n)$. {This demonstrates, for kitten states generated by the OAT protocol, an anticorrelation between the amount of non-stabilizerness, quantified by the SRE, and the amount of many-body Bell correlations, reflected by the value of $\mathcal{E}$.}
{
This trend also persists for the best squeezed state. At $t = t_{\mathrm{best}}$, $Q$ increases only sub-linearly with $N$, signaling weaker Bell correlations than in kitten states. These weaker Bell correlations are accompanied by increased non-stabilizerness, reflected in the logarithmic divergence of $\mathcal{M}_2$ with $N$.
}

Finally, we observe a plateau for the maximal SRE at a given $N$ at intermediate times between kitten states that seem to show superpositions of sheared probability distributions, akin to squeezing.
The maximum value of the SRE scales logarithmically with the size of the system $N$, \cref{fig:kittens}.

\subsection{Comparing magic state resources of spin-squeezed states generated through one- and two-axis twisting}\label{sec:compare_TACT_OAT}

We study the SRE of spin squeezed states generated by one-axis twisting (OAT) and two-axis counter-twisting (TACT) protocols for short time scales $t \sim t_{\rm best}$.
We define the OAT protocol through the unitary operator $\hat{U}(t) = \exp\{-i \chi t \hat{Z}^2/4\}$, with $\hat{Z}= \sum_j \hat{Z}_j$, acting on the initial state $\ket{X}$, where $\chi t_{\rm best} \sim 3^{1/6} N^{-2/3}$ \cite{Sinatra2011}.
We define the TACT protocol through the unitary operator $\hat{U}(t) = \exp\{-i \chi t (\hat{Z}\hat{Y} + \hat{Y}\hat{Z})/4\}$, with $\hat{Y}= \sum_j \hat{Y}_j$, acting on the initial state $\ket{X}$.
While difficult to realize experimentally \cite{Miller2024,Ma2024}, states generated through TACT may reach Heisenberg limited sensitivity $\Delta\phi = \xi/\sqrt{N}  \sim N^{-1}$ in times scales $\chi t_{\rm best} \sim \log(2N) / (2N)$ \cite{Kajtoch2015}.
We verified maximally spin-squeeezed states generated during the TACT protocol show similar scaling with $N$ as in the OAT protocol.

\begin{figure*}
    \centering
    \includegraphics[width=\linewidth]{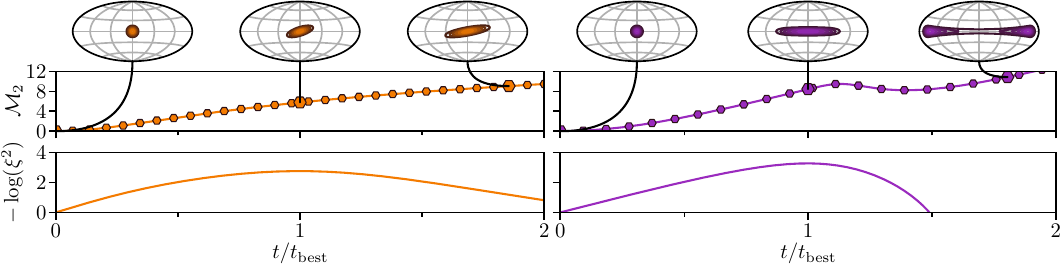}
    \caption{
    Time evolution of SRE, ${\cal M}_2(\ket{\psi(t)})$ (top row), and spin-squeezing parameter $\xi^2$ (bottom row), under OAT (left, orange) and TACT (right, purple) dynamics for $N=100$ spins, on their respectives spin squeezing time scales. 
    Markers indicate exact SRE calculations, while solid lines show results for \cref{eq:SRE_approx}. Insets show the Husimi function $Q(\theta,\phi) = |\bra{\theta, \phi}\ket{\psi(t)}|^2$ at different marked times.
    }
    \label{fig:squeezed_OAT_TACT}
\end{figure*}

\begin{figure}
    \centering
    \includegraphics[width=\linewidth]{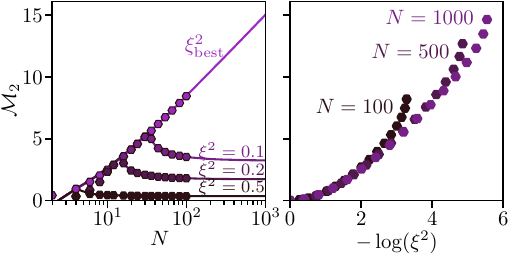}
    \caption{
    Left panel shows scaling of $\mathcal{M}_2$ with $N$ for spin-squeezed states generated by TACT at fixed $\xi^2$. 
    Markers indicate exact SRE calculations, while solid lines show results for \cref{eq:SRE_approx}.
    The label $\xi^2_\mathrm{best}$ corresponds to the best squeezed state with $\xi^2_{\rm best} \propto N^{-1}$. Right panel shows scaling of $\mathcal{M}_2$ with $\xi^2$ for $N=100,500,1000$. 
    }
    \label{fig:fixed_squeezing_TACT}
\end{figure}

We present results for $N=100$ in \cref{fig:squeezed_OAT_TACT}.
In the spin-squeezing time scale, $t\le t_{\rm best}$, both models generate a monotonically increasing SRE while $-\log(\xi^2)$, where $\xi^2$ is the spin squeezing parameter, increases.
At the beginning of the over-squeezed regime, $t > t_{\rm best}$, the SRE continues to increase while $-\log(\xi^2)$ starts to decrease. 
Parallel to the results in the main text for OAT, we demonstrate in \cref{fig:fixed_squeezing_TACT} that under TACT the SRE also saturates to a value fixed by $\xi^2$ when $N\to\infty$. 
As both models offer identical behavior for $t < t_{\rm best}$, we conjecture that spin squeezed states generate SRE according to their metrological advantage measured by $\xi^2$, independently of the protocol used to generate them.

\begin{figure}[t!]
    \centering
    \includegraphics[width=\linewidth]{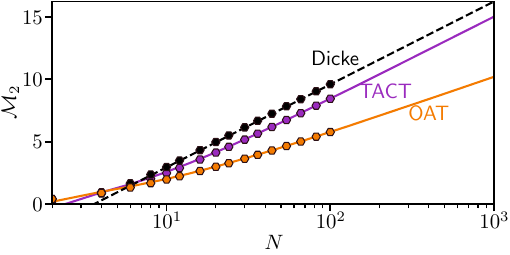}
    \caption{
   SRE scaling with $N$ for the best squeezed state generated with OAT (orange), generated with TACT (purple), and the Dicke state with zero magnetization $\ket{J = N/2, m = 0}$ (black).
    The dashed black line represents \cref{eq:Dicke}.
    Markers indicate exact SRE calculations, while solid lines show results for \cref{eq:SRE_approx}.
    }
    \label{fig:Dicke}
\end{figure}

We show in \cref{fig:Dicke} that TACT provides the same conclusions as OAT for their corresponding maximally squeezed state: the maximally squeezed state generated through TACT shows similar scaling with $N$ to its OAT counterpart.

\subsection{Magic state resources of Dicke states with zero magnetization}

Dicke states can exhibit spin squeezing when their average magnetization is zero,  $\ket{\psi_{\rm D}}=\ket{J=N /2, 0}$, implementing
\emph{twist-and-turn} strategy \cite{Zhang2014,Muessel2015,Carrasco2022, Carrasco2024, Odelli2024}, allowing  to exploit the squeezing and entanglement of these unpolarized states. The spin squeezing parameter can still be well-defined and finite for such states, reflecting the reduced quantum variance in one angular momentum component while others are anti-squeezed \cite{Masson2019}.
{Dicke state in the zero magnetization sector can be seen as
an extreme case of a kitten state with $n=N$, since
$\sum_{j=1}^{N}(-1)^j\ket{\theta= \pi /2, \phi=2\pi j / N} \propto \ket{\psi_{\rm D}}$.
}
 {Evaluation of} the projections of $\ket{\psi_{\rm D}}$ onto the relevant spin coherent states is straightforward,
yielding
$|c_{n,2}^{(X)}| = |c_{n,2}^{(Y)}| = \frac{2}{2^{2J}} {2J \choose J}$,
with  {all} other coefficients vanishing. The binomial coefficient can be approximated using Stirling's Formula as $\lim_{J\to\infty}{2J \choose J} = 2^{2J} / \sqrt{\pi J}$  {resulting in}
\begin{equation}
    \label{eq:Dicke}
    \lim_{N\to\infty} \mathcal{M}_q(\ket{\psi_{\rm D}})
    \simeq  \frac{q}{q-1}\log_2\left(\frac{\pi J}{4}\right) - \frac{1}{q-1},
\end{equation}
{where $J=N/2$.}
This result {indicates} a {logarithmic} scaling SRE $\mathcal M_q\propto \log_2(N)$    {reproducing the} scaling  {of the most} squeezed states generated via OAT dynamics, see
\cref{fig:Dicke}.

The many-body Bell correlator ${\mathcal E}$ for Dicke state $|J=N/2, m\rangle$ has analytical form in the large $N$ limit \cite{Plodzien2025PRA} $
    {\mathcal E} \approx \frac{2}{\pi N},
    $
where $m$ is magnetization. It is maximized for zero magnetization, $m=0$, having the ${\mathcal E} \approx \frac{2}{\pi N}$, $Q \approx N - \log_2 N + \log_2 \frac{2}{\pi}$.

\subsection{Magic state resources and many-body Bell correlations of generalized GHZ states}

We study the SRE and $Q$ of a simple superposition of two spin coherent states as a building block for more complex states involving macroscopically distinct states.
We focus on a superposition of two spin coherent states separated by an arbitrary solid angle in phase space that is appropriately oriented in the XZ-plane, known in a different context as generalized GHZ states~\cite{durEffectiveSizeCertain2002}:

\begin{equation}
	\ket{\psi\left(2\epsilon  \right) } 
    = 
    \frac{1}{\sqrt{K}}\left(
    \ket{\theta=0, \phi=0} + \ket{\theta = 2\epsilon, \phi=0}
    \right)
,\end{equation}
where $K = 2(1 +\cos^{2J}\epsilon)$ \cite{durEffectiveSizeCertain2002}. Without loss of generality, we impose $2\epsilon \in [0, \pi]$.
{For $\epsilon = \pi/2$ it corresponds to GHZ state along $z$-axis.}

\begin{figure}
    \centering
    \includegraphics[width=\linewidth]{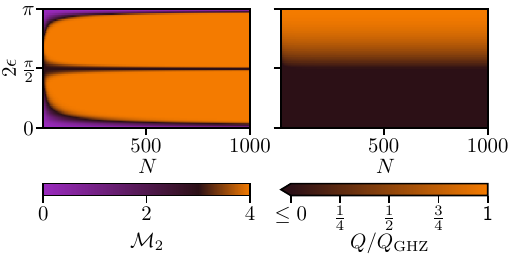}
    \caption{
{
    Approximated SRE (left panel) and many-body Bell correlator $Q$ (right panel) scaling with system size $N$ for generalized GHZ states parametrized by angle $\epsilon$, $\ket{\psi(2\epsilon)}$ for $2\epsilon\in[0,\pi]$. The absolute error with respect to exact calculations of $\mathcal{M}_2$ decays with $N$, and at $N = 100$ is $\lesssim  10^{-2}$.
    $Q$ is normalized with respect to its maximal possible value $Q_{\rm GHZ}=N-2$, with $\hat{J}_+ =\frac{1}{2}(\hat{X} + i\hat{Y})$, i.e. rising operator along the $z$-axis.
    }
    }
   
    \label{fig:GHZ-Dicke}
\end{figure}

We start with SRE analysis. The relevant coefficients for these states are
\begin{align*}
    c_{1,m}^{(X)} =& \frac{2^{-2J+1}}{1+\cos^{2J}\epsilon}\bigg[ (1+(\cos\epsilon + \sin\epsilon)^{2J})^2 \\
    & \pm (1+(\cos\epsilon - \sin\epsilon)^{2J})^2 \bigg],\\
    c_{1,m}^{(Z)} =& \frac{1}{2}\left[
    \frac{\sin^{4J}\epsilon}{(1+\cos^{2J}\epsilon)}
    \pm (1+\cos^{2J}\epsilon)
    \right],\\ 
    c_{2,2}^{(Z)} =& \sin^{2J}\epsilon,
\end{align*}
while all other coefficients vanish or decay with at least $2^{-J}$ for any $\epsilon$.
Taking the $N\to\infty$ limit we obtain
\begin{equation*}
    \lim_{N\to\infty}{\cal M}_q(\ket{\psi(\epsilon)}) 
    \simeq
    \begin{cases}
        0 & 2\epsilon \sim 0,\\
        (2q-1)/(q-1) & 2\epsilon \sim \pi / 2,\\
        0 & 2\epsilon \sim \pi,\\
        2q /(q-1) & \mathrm{else}.
    \end{cases} 
\end{equation*}
Zero SRE is expected for $2\epsilon \sim \{0, \pi\}$, as they correspond to the $\ket{-Z}$ and the regular GHZ state, respectively.
We numerically verified that Generalized GHZ states have a fixed SRE for a sufficiently large $N$, see \cref{fig:GHZ-Dicke}. 

{
We find $Q$ monotonically grows as the state gets closer to the regular GHZ state. 
We may detect many-body Bell correlations for $2\epsilon > \pi /2$, while lower values of $2\epsilon$ yield non-positive $Q$, representative of separable states.
}

\section{Experimental feasibility} 

The SRE can be measured  experimentally without any assumptions on the structure of $\ket{\psi}$ by quantum algorithms that involve Bell measurements requiring access $O(q)$ copies of the state, and for even $q$, additionally the access to the complex conjugate of the state~\cite{Oliviero22, Haug23, Haug24}. For permutationally invariant states, the sample complexity requirements can be alleviated since the full quantum state tomography can be performed with measurement effort $O(N^2)$~\cite{Toth2010}, enabling evaluation of the SRE with a classical postprocessing cost of $O(N^3)$ employing the formula of Ref.~\cite{Passarelli2024}. Our result, Eq.~\eqref{eq:SRE_approx},  provides a \emph{drastic} simplification over these approaches, enabling evaluation of a good approximation for SRE for the permutationally invariant states as a function of only $6$ overlaps $\langle\pm \sigma | \psi \rangle$, \emph{independently} of $N$. These, in general complex, overlaps can be measured in experimentally relevant setting following the interaction-based readout (twisting-echo) scheme~\cite{Davis2016}, on a platform with standard OAT controls~\cite{Riedel2010} and time-reversal metrology demonstrations~\cite{Colombo2022}.
 
In the following, we discuss an experimental protocol allowing measurement of the six complex projections relevant for the SRE determination:
\begin{equation}
a_{\pm\sigma}\equiv \langle \pm\sigma|\psi\rangle = \textrm{Re}
a_{\pm\sigma}+ i \textrm{Im} a_{\pm\sigma},\qquad \sigma\in\{X,  Y, Z\},
\label{eq:goal-a}
\end{equation}
for a permutationally symmetric state $|\psi\rangle$ realized in, e.g., internal-state BEC. The approach follows the interaction-based readout (twisting-echo) protocol~\cite{Davis2016}, implemented on a platform with standard OAT controls~\cite{Riedel2010} and time-reversal metrology demonstrations~\cite{Colombo2022}.
We work in the symmetric $J=N/2$ manifold with
\begin{equation}
\mathbf{\hat{J}}=(\hat J_x, \hat J_y, \hat J_z)=\tfrac12\sum_{i=1}^N\boldsymbol{\hat{\sigma}^{(i)}},\qquad
\hat R_\mu(\vartheta)=e^{-i\vartheta \hat J_\mu}.
\label{eq:collective}
\end{equation}
The state of interest is prepared by one-axis twisting (OAT),
\begin{equation}
\hat U_\chi(t)=e^{-i\chi t\,\hat J_z^2},\qquad |\psi(t)\rangle=\hat U_\chi(t)\,|+\!X\rangle.
\label{eq:OAT}
\end{equation}

Let us define the cardinal coherent states as
\begin{align}
|+\!Z\rangle=&|\!\uparrow\rangle^{\otimes N},\quad
|+\!X\rangle=\hat R_y\!\Big(\tfrac{\pi}{2}\Big)|+\!Z\rangle, \nonumber \\
|&+\!Y\rangle=\hat R_x\!\Big(-\tfrac{\pi}{2}\Big)|+\!Z\rangle.
\label{eq:cardinals}
\end{align}
The opposite directions are obtained by a $\pi$ rotation about any axis orthogonal to the target:
\begin{align}
|-\!Z\rangle=\hat R_x(\pi)|+\!Z\rangle=R_y(\pi)|+\!Z\rangle, \nonumber \\
|-\!X\rangle=\hat R_y(\pi)|+\!X\rangle=R_z(\pi)|+\!X\rangle, \nonumber \\
|-\!Y\rangle=\hat R_x(\pi)|+\!Y\rangle=R_z(\pi)|+\!Y\rangle.
\label{eq:minus}
\end{align}

For detection we use the top-Dicke projector $\hat \Pi_J=|+\!Z\rangle\!\langle+\!Z|$ together with $\hat M_{+X}=\hat R_y(-\tfrac{\pi}{2})$, $\hat M_{+Y}= \hat R_x(\tfrac{\pi}{2})$, $\hat M_{+Z}=\mathbb{\hat I}$ so that
\begin{equation}
\hat \Pi_J^{(\sigma)}\equiv \hat M_{+\sigma}^\dagger \hat \Pi_J \hat M_{+\sigma}=|+\!\sigma\rangle\!\langle+\!\sigma|.
\label{eq:proj-map}
\end{equation}

To experimentally probe overlaps we apply a small analysis rotation confined to the tangent plane at $+\sigma$,
\begin{equation}
\hat R_{\sigma}^{\mathrm{(tan)}}(\theta,\phi)
=\exp\!\Big[-i\theta\big(\cos\phi\,\hat J_\beta+\sin\phi\,\hat J_\gamma\big)\Big],
\label{eq:Rtan}
\end{equation}
which produces a first-order displacement, where we fix the sign convention by taking $(\beta,\gamma,\sigma)$ to be right–handed, i.e.
$[\hat J_\beta,\hat J_\gamma]=i\hat J_\sigma$. For axis $\sigma=X$ (others are relabelings), the twist–kick–untwist with a final map $+X\!\to\!+Z$ yields
\begin{equation}
|\psi_{\rm out}^{(X)}(\phi)\rangle
= \hat M_{+X}\,\hat U_\chi^\dagger(t)\,R_X^{\mathrm{(tan)}}(\theta,\phi)\,\hat U_\chi(t)\,|+\!X\rangle,
\label{eq:psi-out}
\end{equation}
with amplitude and probability for all spins-up configuration $ P_J^{(X)}(\phi)=|A_X(\phi)|^2$ where
\begin{align}
A_X(\phi)=\langle+\!Z|\psi_{\rm out}^{(X)}(\phi)\rangle 
=\langle+\!X|\hat U_\chi^\dagger \hat R_X^{\mathrm{(tan)}}(\theta,\phi)\hat U_\chi|+\!X\rangle.
\label{eq:AX}
\end{align}
Linearizing Eq.\eqref{eq:Rtan} around small angle $\theta\ll 1$ gives
\begin{align}
\hat K_X(\phi)&=\cos\phi\,\hat J_Y+\sin\phi\,\hat J_Z, \nonumber \\
\hat R_X^{\mathrm{(tan)}}(\theta,\phi)&=\mathbb{I}-i\theta \hat K_X(\phi)+\mathcal O(\theta^2),
\label{eq:KX}
\end{align}
and we obtain
\begin{equation}
P_J^{(X)}(\phi)=1+2\theta\,\mathrm{Im}\langle\psi|\hat K_X(\phi)|\psi\rangle+\mathcal O(\theta^2).
\label{eq:P-lin}
\end{equation}
Next, by inserting identity resolution $\mathbb{\hat{I}}=|+\!X\rangle\!\langle+\!X|+\hat Q_X$ to the right of $\hat K_X(\phi)$, with transverse projector $ \hat Q_X=\mathbb{I}-|+\!X\rangle\!\langle+\!X|$ and $\hat Q_X|+\!X\rangle=0$, we get
\begin{align}
\langle\psi|\hat K_X(\phi&)|\psi\rangle
=\langle\psi|\hat K_X(\phi)\big(|+\!X\rangle\!\langle+\!X|+\hat Q_X\big)|\psi\rangle \nonumber\\
&=\underbrace{\langle+\!X|\hat K_X(\phi)|\psi\rangle}_{c_X(\phi)}\,
\underbrace{\langle+\!X|\psi\rangle}_{a_{+X}}
+\underbrace{\langle\psi|\hat K_X(\phi)\hat Q_X|\psi\rangle}_{\delta_X(\phi)}.
\label{eq:split}
\end{align}
Since $c_X(\phi)=c_Y\cos\phi+c_Z\sin\phi$ with $c_Y=\langle+\!X|J_Y|\psi\rangle$, $c_Z=\langle+\!X|J_Z|\psi\rangle$, the $\mathcal O(\theta)$ fringe is a sinusoid,
\begin{equation}
P_J^{(X)}(\phi)=A_X+B_X\!\left[\mathrm{Re}\,a_{+X}\cos\phi-\mathrm{Im}\,a_{+X}\sin\phi\right]+\mathcal O(\theta^2),
\label{eq:AB}
\end{equation}
where $A_X$ absorbs $\phi$-independent offsets and $B_X\propto\theta$ is a real gain. To determine $A_x$ and $B_X$ we prepare 
calibration at $t=0$ ($|\psi\rangle=|+\!X\rangle$, $a_{+X}=1$)
\begin{align}
A_X=\tfrac12\!\left[P_{\rm cal}^{(X)}(0)+P_{\rm cal}^{(X)}(\pi)\right], \nonumber \\
B_X=\tfrac12\!\left[P_{\rm cal}^{(X)}(0)-P_{\rm cal}^{(X)}(\pi)\right].
\label{eq:ABcal}
\end{align}
Thus we obtain the quadratures from two phases,
\begin{equation}
\mathrm{Re}\,a_{+\sigma}\approx\frac{P_J^{(\sigma)}(0)-A_\sigma}{B_\sigma},\quad
\mathrm{Im}\,a_{+\sigma}\approx-\,\frac{P_J^{(\sigma)}(\tfrac{\pi}{2})-A_\sigma}{B_\sigma},
\label{eq:ReIm}
\end{equation}
 
Generalization to all projectors $a_{\pm\sigma}$ is straigtforward.
Let $(\beta,\gamma,\sigma)$ ; for $\sigma=X$: $(\beta,\gamma)=(Y,Z)$ and cyclic permutations otherwise. For each sign $s\in\{+,-\}$ define the detection map $M_{s\sigma}$ that sends $|s\sigma\rangle$ to $|+\!Z\rangle$. A convenient construction is
\begin{equation}
\hat M_{+\sigma}=\hat M_{+X},\hat M_{+Y}, \hat M_{+Z},\qquad
\hat M_{-\sigma}=\hat M_{+\sigma}\,\hat R_{\perp}^{(\sigma)}(\pi),
\label{eq:Msalpha}
\end{equation}
where $\hat R_{\perp}^{(\sigma)}(\pi)$ is \emph{any} $\pi$ rotation about an axis orthogonal to $\sigma$ (so $\hat R_{\perp}^{(\sigma)}(\pi)\,|-\sigma\rangle=|+\sigma\rangle$). The tangent–plane analysis kick for the pole $s\sigma$ can be realized either explicitly or by a phase shift:
\begin{equation}
\hat R_{s\sigma}^{\mathrm{(tan)}}(\theta,\phi)
=\exp\!\Big[-i\theta\big(\cos\phi\,\hat J_{\beta}^{(s)}+\sin\phi\,\hat J_{\gamma}^{(s)}\big)\Big],
\label{eq:Rtan-sign}
\end{equation}
with $\hat J_{\beta,\gamma}^{(s)}=s\,\hat J_{\beta,\gamma}$.

The \emph{exact} twist–kick–untwist sequence for a general projector is
\begin{equation}
|\psi_{\rm out}^{(s,\sigma)}(\phi)\rangle
= \hat M_{s\sigma}\,\hat U_\chi^\dagger(t)\,\hat R_{s\sigma}^{\mathrm{(tan)}}(\theta,\phi)\,\hat U_\chi(t)\,|+\!X\rangle,
\label{eq:psiout-gen}
\end{equation}
and the measured probability of all spins-up configuration is $P_J^{(s,\sigma)}(\phi)=|\langle+\!Z|\psi_{\rm out}^{(s,\sigma)}(\phi)\rangle|^2
=|\langle s\sigma|U_\chi^\dagger \hat R_{s\sigma}^{\mathrm{(tan)}}(\theta,\phi)\hat U_\chi|+\!X\rangle|^2$.
Linearizing in $\theta$ and inserting $\mathbb{\hat{I}}=|s\sigma\rangle\langle s\sigma|+\hat Q_{s\sigma}$ to the right of the generator $\hat K_{s\sigma}(\phi)=\cos\phi\,\hat J_{\beta}^{(s)}+\sin\phi\,\hat J_{\gamma}^{(s)}$ yields the same sinusoidal form as Eq.~\eqref{eq:AB} but with $a_{+X}\to a_{s\sigma}$:
\begin{equation}
P_J^{(s,\sigma)}(\phi)=A_{s\sigma}+B_{s\sigma}\!\left[\mathrm{Re}\,a_{s\sigma}\cos\phi-\mathrm{Im}\,a_{s\sigma}\sin\phi\right]+\mathcal O(\theta^2).
\label{eq:AB-gen}
\end{equation}
Accordingly, the two–phase estimates are identical in form,
\begin{equation}
\mathrm{Re}\,a_{s\sigma}\approx\frac{P_J^{(s,\sigma)}(0)-A_{s\sigma}}{B_{s\sigma}},\quad
\mathrm{Im}\,a_{s\sigma}\approx-\,\frac{P_J^{(s,\sigma)}(\tfrac{\pi}{2})-A_{s\sigma}}{B_{s\sigma}},
\label{eq:ReIm-gen}
\end{equation}
with $(A_{s\sigma},B_{s\sigma})$ obtained from a $t{=}0$ calibration for that $(s,\sigma)$.

\section{Conclusions}
We study the non-stabilizerness of permutationally symmetric quantum many-body states and present a compact analytic approximation for the SRE in the large-system limit. The formula depends on only a constant number of expectation values, substantially simplifying experimental quantification of non-stabilizerness and providing a practical tool for analytical evaluation in the permutationally invariant sector. We leverage our formula to understand the non-stabilizerness in spin-squeezing protocols.

Focusing on the OAT protocol, we show that spin-squeezing protocols inherently generate non-stabilizerness. 
Best-squeezed states produced by OAT exhibit an SRE which grows logarithmically with the system size, while states with a fixed value of the spin squeezing parameter, which quantifies metrological usefulness, exhibit an SRE independent of the system size.
Another family of metrologically useful states generated by OAT—so-called kitten states, superpositions of rotated GHZ states—exhibits an SRE that remains independent of system size. The limited non-stabilizerness of the kitten states is anticorrelated with their many-body Bell correlations: while the SRE increases with the number of heads $n$ in the kitten state, the corresponding Bell correlations decrease.
We also observe a relation between the scaling of the SRE with system size and the robustness of the quantum state. 
Spin-squeezed states, which are robust against perturbations, provide increasing SRE according to their metrological usefulness.
In contrast, the GHZ state, while achieving the Heisenberg limit, is highly vulnerable to perturbations and has vanishing SRE.

Our observations for the behavior of the SRE in the squeezed states of the OAT protocol are corroborated by results for the two-axis countertwisting (TACT) protocol \cite{Miller2024,Ma2024} and for Dicke states with zero magnetization. In both cases, the SRE increases logarithmically with system size, as in the best-squeezed OAT states. 
The best-squeezed state in TACT and Dicke states, similarly to the best-squeezed state in OAT,  feature weaker many-body Bell correlations than the kitten states, further exemplifying the anticorrelation between SRE and many-body Bell correlations.
These findings point to broader connections between non-stabilizerness, quantum metrology, and multipartite quantum correlations that we leave for further investigations.

\section*{Acknowledgments}
P.S. acknowledges insightful discussions on magic-state resources with E. Tirrito, X. Turkeshi, and members of the Quantic group at BSC.
The work of T.H.-Y. and J.Z. was funded by the National Science Centre, Poland, project 2021/03/Y/ST2/00186 within the QuantERA II Programme that has received funding from the European Union Horizon 2020 research and innovation programme under Grant agreement No 101017733.  
M.P. acknowledges support from:
European Research Council AdG NOQIA; MCIN/AEI (PGC2018-0910.13039/501100011033, CEX2019-000910-S/10.13039/501100011033, Plan National FIDEUA PID2019-106901GB-I00, Plan National STAMEENA PID2022-139099NB, I00, project funded by MCIN/AEI/10.13039/501100011033 and by the “European Union NextGenerationEU/PRTR" (PRTR-C17.I1), FPI); QUANTERA DYNAMITE PCI2022-132919, QuantERA II Programme co-funded by European Union’s Horizon 2020 program under Grant Agreement No 101017733; Ministry for Digital Transformation and of Civil Service of the Spanish Government through the QUANTUM ENIA project call - Quantum Spain project, and by the European Union through the Recovery, Transformation and Resilience Plan - NextGenerationEU within the framework of the Digital Spain 2026 Agenda; MICIU/AEI/10.13039/501100011033 and EU (PCI2025-163167);
Fundació Cellex; Fundació Mir-Puig; Generalitat de Catalunya (European Social Fund FEDER and CERCA program; M.P.  acknowledges RES resources provided by Barcelona Supercomputing Center in Marenostrum 5 to FI-2025-1-0043.
P.S. acknowledges fellowship within the “Generación D” initiative, Red.es, Ministerio para la Transformación Digital y de la Función Pública, for talent attraction (C005/24-ED CV1), funded by the European Union NextGenerationEU funds, through PRTR.

\appendix

{

\section{Pauli strings in the overcomplete spin coherent basis}
\label{sup:pauli_spectra}

Metrologically useful states like spin squeezed states, kitten states, and GHZ states are permutation invariant. We must project the relevant observables in the permutation invariant sector to efficiently study their non-stabilizerness properties. As we are particularly interested in the stabilizer Rényi entropy (SRE) as our measure of non-stabilizerness, we must first compute its constituents, the so-called Pauli strings.

We project the Pauli string operators $\hat{P}\equiv \otimes_{k=1}^N \hat{P}_k$, where $\hat{P}_k \in \left\{ \hat{X}_k, \hat{Y}_k, \hat{Z}_k, \hat{\mathbb{I}}_k \right\} $ is one of the Pauli and identity operators acting on qubit $k$, on the overcomplete spin coherent state basis as
\begin{multline}
    \label{eq:pauli_arbitrary_sup}
        \hat{\mathbb{I}}_{J}\hat{P}\hat{\mathbb{I}}_{J} 
		= \left( \frac{2J+1}{4\pi} \right)^2 \int \int d\vb*{\Omega_i} d\vb*{\Omega_j} 
		 \ket{\theta_i, \phi_i} \bra{\theta_j, \phi_j} 
	\\
	\cdot 
	\bra{\theta_i, \phi_i} \hat{P} \ket{\theta_j, \phi_j}
.\end{multline}

While spin coherent states are typically expressed in the angular momentum eigenbasis for total spin $J = N /2$, also known as the Dicke basis,
here we choose the spin coherent representation for convenience when dealing with Pauli operators.
We can choose an arbitrary reference spin coherent state, for instance
$\ket{\theta=0, \phi=0} = \ket{J, J} = \otimes_{j=1}^N \ket{1}_j$.
and write any spin coherent state as the action of two orthogonal collective rotations on it like
\begin{equation*}
	\ket{\theta, \phi} = e^{-i\theta \hat{S}_y} e^{-i \phi \hat{S}_z} \bigotimes_{k=1}^N\ket{1}_k
,\end{equation*}
where $\hat{S}_y = \sum_j \hat{Y}_j /2, \hat{S}_z = \sum_j \hat{Z}_j /2$.
Given an operator $\hat{A}=\sum_l \otimes_k \hat{A}_{k,l}$ that acts on each subsystem $k$ with the configuration $l$, 
we can express its matrix elements in the spin coherent basis as
\begin{widetext}
\begin{align*}
	\bra{\theta_i, \phi_i}\hat{A}\ket{\theta_j, \phi_j} 
	&= \bigotimes_{k=1}^N \bra{1}_k e^{i\theta_i \hat{S}_y} e^{i \phi_i \hat{S}_z} \hat{A} e^{-i\theta_j \hat{S}_y} e^{-i \phi_j \hat{S}_z} \bigotimes_{k'=1}^N \ket{1}_{k'}\\
	&= \sum_{l} \bigotimes_{k=1}^N \left( \bra{1}_k e^{i\theta_i /2\hat{Y}_k} e^{i \phi_i /2 \hat{Z}_k} \hat{A}_{k,l} e^{-i\theta_j /2 \hat{Y}_k} e^{-i \phi_j /2 \hat{Z}_k} \ket{1}_k \right) 
.\end{align*}
The matrix elements of a Pauli string are then given by
\begin{equation*}
	\bra{\theta_i, \phi_i}\hat{P}\ket{\theta_j, \phi_j} =
	\bigotimes_{k=1}^N \left( \bra{1}_k e^{i\theta_i /2\hat{Y}_k} e^{i \phi_i /2 \hat{Z}_k} \hat{P}_k e^{-i\theta_j /2 \hat{Y}_k} e^{-i \phi_j /2 \hat{Z}_k} \ket{1}_k \right) 
.\end{equation*}
\end{widetext}

Each term in the tensor product only depends on the particular choice of Pauli operator $\hat{P}_k$. 
We can rearange the expression as the product of $N_\sigma$ identical terms for each of the four available operators $\hat{\sigma} \in \{\hat{X}, \hat{Y}, \hat{Z}, \hat{\mathbb{I}}\}$.
Then,
\begin{equation}\label{eq:Pauli_matrix_ij}
	\bra{\theta_i, \phi_i}\hat{P}\ket{\theta_j, \phi_j} =
	(\alpha_{i,j})^{N_X} (\beta_{i,j})^{N_Y} (\gamma_{i,j})^{N_Z} (\kappa_{i,j})^{N_I}
,\end{equation}
where we identify the coefficients with
\begin{align}
	\label{eq:a_raw}
	\alpha_{i,j} &= 
	\bra{1} e^{i\theta_i /2\hat{Y}} e^{i \phi_i /2 \hat{Z}} \hat{X} e^{-i\theta_j /2 \hat{Y}} e^{-i \phi_j /2 \hat{Z}} \ket{1}
	,\\
	\label{eq:b_raw}
	\beta_{i,j} &= 
	\bra{1} e^{i\theta_i /2\hat{Y}} e^{i \phi_i /2 \hat{Z}} \hat{Y} e^{-i\theta_j /2 \hat{Y}} e^{-i \phi_j /2 \hat{Z}} \ket{1}
	,\\
	\label{eq:c_raw}
	\gamma_{i,j} &= 
	\bra{1} e^{i\theta_i /2\hat{Y}} e^{i \phi_i /2 \hat{Z}} \hat{Z} e^{-i\theta_j /2 \hat{Y}} e^{-i \phi_j /2 \hat{Z}} \ket{1}
	,\\
	\label{eq:d_raw}
	\kappa_{i,j} &= 
	\bra{1} e^{i\theta_i /2\hat{Y}} e^{i \phi_i /2 \hat{Z}} \hat{\mathbb{I}} e^{-i\theta_j /2 \hat{Y}} e^{-i \phi_j /2 \hat{Z}} \ket{1}
	.
\end{align}

Since for any operator $\hat{B}$ for which $\hat{B}^2 = \hat{\mathbb{I}}$ we can write $e^{i\theta \hat{B}} = \cos \theta \hat{\mathbb{I}} + i\sin\theta \hat{B}$, we can express the rotations as 
\begin{equation*}
    \begin{split}
    	e^{i\theta /2\hat{Y}} e^{i \phi /2 \hat{Z}} 
        =& \left( \cos(\theta/2)\hat{\mathbb{I}} + i\sin(\theta /2)\hat{Y} \right)\\
    	&\cdot\left( \cos(\phi/2)\hat{\mathbb{I}} + i\sin(\phi /2)\hat{Z} \right).
    \end{split}
\end{equation*}

The algebra of the Pauli operators allow us to reduce their products into instances of a single operator with relations like $\hat{X}^2=\hat{Y}^2=\hat{Z}^2 = -i \hat{X}\hat{Y}\hat{Z} = \hat{\mathbb{I}}$, $\hat{X}\hat{Y} = i\hat{Z}$ and so on. 
This allows us to expand the products on each coefficient and reduce to simple expressions using trigonometric relations.
The calculation is tedious, but straigthforward.

The final expressions for the coefficients after taking into account $\bra{1}\hat{X}\ket{1} = \bra{1}\hat{Y}\ket{1} = 0$ and  $\bra{1}\hat{Z}\ket{1} = \bra{1}\hat{\mathbb{I}}\ket{1} = 1$ are 
\begin{align}
    \begin{split}
	\label{eq:a}
	\alpha_{i,j} =&
	\cos\left( \frac{\phi_i + \phi_j}{2} \right) \sin\left( \frac{\theta_i + \theta_j}{2}  \right)
	\\
    &- i \sin\left( \frac{\phi_i + \phi_j}{2} \right) \sin\left( \frac{\theta_i - \theta_j}{2}  \right)
	,\end{split}\\
    \begin{split}
	\label{eq:b}
	\beta_{i,j} =&
	\sin\left( \frac{\phi_i + \phi_j}{2} \right) \sin\left( \frac{\theta_i + \theta_j}{2}  \right)
	\\
    &+ i \cos\left( \frac{\phi_i + \phi_j}{2} \right) \sin\left( \frac{\theta_i - \theta_j}{2}  \right)
	,\end{split}\\
    \begin{split}
	\label{eq:c}
	\gamma_{i,j} =&
	\cos\left( \frac{\phi_i - \phi_j}{2} \right) \cos\left( \frac{\theta_i + \theta_j}{2}  \right)
	\\
    &+ i \sin\left( \frac{\phi_i - \phi_j}{2} \right) \cos\left( \frac{\theta_i - \theta_j}{2}  \right)
	,\end{split}\\
    \begin{split}
	\label{eq:d}
	\kappa_{i,j} =&
	\cos\left( \frac{\phi_i - \phi_j}{2} \right) \cos\left( \frac{\theta_i - \theta_j}{2}  \right)
	\\
    &+ i \sin\left( \frac{\phi_i - \phi_j}{2} \right) \cos\left( \frac{\theta_i + \theta_j}{2}  \right)
	.\end{split}
\end{align}

\section{System size independent Pauli string matrix elements}
\label{sup:pauli_solutions_1100}

The analytical integration of \cref{eq:pauli_arbitrary_sup} is a difficult task that might yield unsatisfactory solutions lacking a closed form. 
As a good approximation is sufficient for our purposes, we look for possible simplifications that return a limited number of closed form matrix elements in the spin coherent basis. 
A good starting point is to choose constraints that account for the largest contributions in the integration domain. 
Since any Pauli string has eigenvalues $\pm 1$ and \cref{eq:Pauli_matrix_ij} indicates an exponential dependence in the system size coefficients $N_X, N_Y, N_Z, N_I$ (where $N_X+N_Y+N_Z+N_I = N$), we predict the maximal contributions to the integrals come from system size independent terms.
Other naively good constraint choices may limit, at least partially, the influence of the system size on the integrands,
but might also result in additional complications when computing the SRE. 
We find our proposed constraints are sufficient for our interests and expect to be useful in other contexts.

We are interested in particular terms 
of \cref{eq:pauli_arbitrary_sup} that do not change with the system size, so we must find values of the coefficients 
$\alpha_{i,j}$, $\beta_{i,j}$, $\gamma_{i,j}$, $\kappa_{i,j}$ 
described in \crefrange{eq:a}{eq:d} 
such that the combined result is independent of the exponents 
$N_X$, $N_Y$, $N_Z$, $N_I$. As these coefficients are complex, it is convenient to instead study
$|\alpha_{i,j}|^2$, $|\beta_{i,j}|^2$, $|\gamma_{i,j}|^2$, $|\kappa_{i,j}|^2$.
It is clear from inspection of 
\crefrange{eq:a}{eq:d} 
that  
\begin{equation}\label{eq:coeff_mod_condition}
	0 \le |c|^2 \le 1; \forall c \in \{\alpha_{i,j}, \beta_{i,j}, \gamma_{i,j}, \kappa_{i,j}\}.
\end{equation}

We express these coefficients moduli explicitly as
\begin{align*}
    \begin{split}
    |\alpha_{i,j}|^2 
    =& \sin^{2}{\left(\frac{\phi_i+\phi_j}{2} \right)} \sin^{2}{\left(\frac{\theta_i-\theta_j}{2} \right)}\\
    &+ \sin^{2}{\left(\frac{\theta_i+\theta_j}{2} \right)} \cos^{2}{\left(\frac{\phi_i+\phi_j}{2} \right)}
	,\end{split}\\
    \begin{split}
	|\beta_{i,j}|^2 
    =& \sin^{2}{\left(\frac{\phi_i+\phi_j}{2} \right)} \sin^{2}{\left(\frac{\theta_i+\theta_j}{2} \right)} \\
	&+ \sin^{2}{\left(\frac{\theta_i-\theta_j}{2} \right)} \cos^{2}{\left(\frac{\phi_i+\phi_j}{2} \right)}
	,\end{split}\\
    \begin{split}
	|\gamma_{i,j}|^2 
    =& \sin^{2}{\left(\frac{\phi_i-\phi_j}{2} \right)} \cos^{2}{\left(\frac{\theta_i-\theta_j}{2} \right)} \\
    &+ \cos^{2}{\left(\frac{\phi_i-\phi_j}{2} \right)} \cos^{2}{\left(\frac{\theta_i+\theta_j}{2} \right)}
	,\end{split}\\
    \begin{split}
	|\kappa_{i,j}|^2 
    =& \sin^{2}{\left(\frac{\phi_i-\phi_j}{2} \right)} \cos^{2}{\left(\frac{\theta_i+\theta_j}{2} \right)} \\
    &+ \cos^{2}{\left(\frac{\phi_i-\phi_j}{2} \right)} \cos^{2}{\left(\frac{\theta_i-\theta_j}{2} \right)}
	.\end{split}
\end{align*}

Although we can introduce some symmetry considerations, solving this system of equations can become cumbersome since the trigonometric functions will yield coupled periodic solutions which can be difficult to simplify.
An alternative approach is to substitute them by auxiliary variables 
\begin{align*}
	a &= \cos^2\left( \frac{\phi_i+\phi_j}{2} \right),\\ 
	b &= \cos^2\left( \frac{\phi_i-\phi_j}{2} \right),\\ 
	c &= \cos^2\left( \frac{\theta_i+\theta_j}{2} \right),\\ 
	d &= \cos^2\left( \frac{\theta_i-\theta_j}{2} \right), 
\end{align*}
for which we firstly find unbound solutions for ${a,b,c,d}\in\mathcal{R}$ and at the end of our procedure we impose the domain ${a,b,c,d}\in [0, 1]$. With this substitution we obtain

\begin{align}
	\label{eq:a1}
	|\alpha_{i,j}|^2 =&
	a \left(1 - c\right) + \left(1 - a\right) \left(1 - d\right)
	,\\
	\label{eq:b1}
	|\beta_{i,j}|^2 =& 
	a \left(1 - d\right) + \left(1 - a\right) \left(1 - c\right)
	,\\
	\label{eq:c1}
	|\gamma_{i,j}|^2 =& 
	b c + \left(1 - b\right) d
	,\\
	\label{eq:d1}
	|\kappa_{i,j}|^2 =& 
	b d + \left(1 - b\right) c
	.
\end{align}

From this parametrization it becomes obvious that 
\begin{equation}\label{eq:pauli_constraint}
		|\alpha_{i,j}|^2 + |\beta_{i,j}|^2  + |\gamma_{i,j}|^2 + |\kappa_{i,j}|^2 = 2.
\end{equation}

\Cref{eq:coeff_mod_condition} and \cref{eq:pauli_constraint} tell us that there is only one way for a term in \cref{eq:pauli_arbitrary_sup} to be system size independent: 
two of the coefficients are unitary while the other two vanish. 
In other words, 
$x_1 = x_2 = 1, x_3=x_4 = 0; \forall \{x_1, x_2, x_3, x_4\} \subseteq \mathcal{P}\left(|\alpha_{i,j}|^2, |\beta_{i,j}|^2, |\gamma_{i,j}|^2, |\kappa_{i,j}|^2\right)$,
where $\mathcal{P}$ indicates all possible permutations of the set elements.

We can now  proceed to solve our system of equations under the constraints we just found for each possible combination of terms and then recover the domain of the auxiliary variables $a, b, c, d$ to see which terms stay system indepedent in \cref{eq:pauli_arbitrary_sup}.
For obtaining the final results in terms of $\phi_i, \phi_j, \theta_i, \theta_j$ it will be convenient to keep in mind the following inverse cosine relations:
\begin{align*}
	\arccos\left( 0 \right) =& \frac{\pi}{2}, \\ 
	\arccos\left(  1 \right) =& 0, \\  
	\arccos\left( - 1 \right) =& \pi 
,\end{align*}
which we use to retrieve
\begin{align*}
	\phi_i = a' + b', \\
	\phi_j = a' - b', \\
	\theta_i = c' + d', \\
	\theta_j = c' - d',
\end{align*}
where in $a' = \arccos(\pm\sqrt{a}), b' = \arccos(\pm\sqrt{b}), c' = \arccos(\pm\sqrt{c}), d' = \arccos(\pm\sqrt{d})$ we have to make a sign choice to retrieve every possible solution.

Although the following analysis could be simplified through symmetry arguments, we will explore all possible solutions for the sake of completeness.

\begin{itemize}
	\item {$|\alpha_{i,j}|=|\beta_{i,j}|=1$, $\gamma_{i,j} = \kappa_{i,j} = 0$.}
	
		Solving \cref{eq:a1,eq:b1,eq:c1,eq:d1} under these constrains yields two solutions.

		The first solution, $c = d = 0$, 
		implies $\theta_i \in \{0, \pi\}, \theta_j = \theta_i + \pi$.
		As these values correspond to the poles of the spherical coordinates, we set $\phi_i = \phi_j = 0$.
		This solution is only valid for 
		\begin{align*}
			\bra{0, 0} \hat{P} \ket{\pi, 0} 
			&= \bra{Z} \hat{P} \ket{- Z}, \\
			\bra{\pi, 0} \hat{P} \ket{0, 0} 
			&= \bra{-Z} \hat{P} \ket{Z}.
		\end{align*}

		Explicitly solving $\alpha_{i,j}^{N_X} \beta_{i,j}^{N_Y}$ for each combination of spherical coordinates yields
		\[
			\bra{\pm Z} \hat{P} \ket{\mp Z}
			= (\pm i)^{N_Y} \delta_{N_Z, 0}\delta_{N_I, 0}
		.\] 
	
		The second solution, $a = b = \frac{1}{2}, c = -d$, 
		is identical to the previous solution with unnecessary constraints on $a, b$; since $c, d\in[0,1]$ implies $c = -d = 0$.

	\item {$|\alpha_{i,j}|=|\gamma_{i,j}|=1$, $\beta_{i,j} = \kappa_{i,j} = 0$.}

		Solving \cref{eq:a1,eq:b1,eq:c1,eq:d1} under these constrains yields the solution
		$a = 1 - b, c = \frac{b}{2b - 1}, d = \frac{b-1}{2b-1}$.  
		By the domain constraint $b \in [0,1]$ it can be found that
		$(c,d) \in \{(0,1), (1,0)\}$, where the first result happens for $b=0$ and the second for $b=1$.
		Thus, we obtain two proper solutions: $a = 1, b = 0, c = 0, d = 1$ and $a = 0, b = 1, c = 1, d = 0$.
	
		By retrieving all possible solutions for, for instance, $\phi_i, \phi_j$ given $a=1, b = 0$ we obtain  
		$(\phi_i, \phi_j) = (\pm \pi /2, \mp \pi/2)$ while for $a=0, b=1$ we obtain
		$(\phi_i, \phi_j) = (\pm \pi /2, \pm \pi/2)$.
		We will obtain identital results for $\theta_i, \theta_j$, but with opposite parity for each solution. 
		The first solution is only valid for 
		\begin{align*}
			\bra{\frac{\pi}{2}, \pm\frac{\pi}{2}} \hat{P} \ket{\frac{\pi}{2}, \mp \frac{\pi}{2}} 
			&= \bra{-\frac{\pi}{2}, \pm\frac{\pi}{2}} \hat{P} \ket{-\frac{\pi}{2}, \mp \frac{\pi}{2}} \\
		&= \bra{\pm Y} \hat{P} \ket{\mp Y}
		,\end{align*}
		while the second solution is only valid for
		\begin{align*}
			\bra{\frac{\pi}{2}, \pm\frac{\pi}{2}} \hat{P} \ket{-\frac{\pi}{2}, \pm \frac{\pi}{2}} 
			&= \bra{-\frac{\pi}{2}, \pm\frac{\pi}{2}} \hat{P} \ket{\frac{\pi}{2}, \pm \frac{\pi}{2}} \\
			&= \bra{\pm Y} \hat{P} \ket{\mp Y}
		.\end{align*}

		The solutions can be explicitly expressed as
		\begin{align*}
			\bra{\pm Y} \hat{P} \ket{\mp Y}
			= (\pm i)^{N_Z} \delta_{N_Y, 0}\delta_{N_I, 0}
		.\end{align*}

	\item {$|\alpha_{i,j}|=|\kappa_{i,j}|=1$, $\beta_{i,j} = \gamma_{i,j} = 0$.}
		
		Solving \cref{eq:a1,eq:b1,eq:c1,eq:d1} under these constrains yields the solution
		$a = b, c = \frac{b-1}{2b - 1}, d = \frac{b}{2b-1}$.  
		By the domain constraint $b \in [0,1]$ it can be found that
		$(c,d) \in \{(1,0), (0,1)\}$, where the first result happens for $b=0$ and the second for $b=1$.
		Thus, we obtain two proper solutions: $a = 0, b = 0, c = 1, d = 0$ and $a = 1, b = 1, c = 0, d = 1$.
		From previous results we know that $c=1, d=0$ yields
		$(\theta_i, \theta_j) = (\pm \pi /2, \mp \pi/2)$ while $c=0, d=1$ yields
		$(\theta_i, \theta_j) = (\pm \pi /2, \pm \pi/2)$.
		Meanwhile, $a=b=0$ yields
		$(\phi_i, \phi_j) = (\pi, 0)$ while $a=b=1$ yields
		$(\phi_i, \phi_j) \in \{(0, 0), (\pi, \pi)\}$.
		The first solution is only valid for 
		\begin{align*}
			\bra{\pm\frac{\pi}{2}, \pi} \hat{P} \ket{\mp\frac{\pi}{2}, 0} 
			&= \bra{\pm\frac{\pi}{2}, 0} \hat{P} \ket{\mp\frac{\pi}{2}, \pi} \\
			&= \bra{\pm X} \hat{P} \ket{\pm X}
		,\end{align*}
		while the second solution is only valid for
		\begin{align*}
			\bra{\pm\frac{\pi}{2}, 0} \hat{P} \ket{\pm\frac{\pi}{2}, 0} 
			&= \bra{\pm\frac{\pi}{2}, \pi} \hat{P} \ket{\pm\frac{\pi}{2}, \pi} \\
			&= \bra{\pm X} \hat{P} \ket{\pm X}
		.\end{align*}
	
		The solutions can be explicitly expressed as
		\[
			\bra{\pm X} \hat{P} \ket{\pm X}
			= (\pm 1)^{N_X} \delta_{N_Y, 0}\delta_{N_Z, 0}
		.\]

	\item {$|\beta_{i,j}|=|\gamma_{i,j}|=1$, $\alpha_{i,j} = \kappa_{i,j} = 0$.}
		
		Solving \cref{eq:a1,eq:b1,eq:c1,eq:d1} under these constrains yields the solution
		$a = b, c = \frac{b}{2b - 1}, d = \frac{b-1}{2b-1}$.  
		By the domain constraint $b \in [0,1]$ it can be found that
		$(c,d) \in \{(0,1), (1,0)\}$, where the first result happens for $b=0$ and the second for $b=1$.
		Thus, we obtain two proper solutions: $a = 0, b = 0, c = 0, d = 1$ and $a = 1, b = 1, c = 1, d = 0$.
		From previous results we know 
		the first solution is only valid for 
		\begin{align*}
			\bra{\pm\frac{\pi}{2}, \pi} \hat{P} \ket{\pm\frac{\pi}{2}, 0} 
			&= \bra{\pm\frac{\pi}{2}, 0} \hat{P} \ket{\mp\frac{\pi}{2}, \pi} \\
			&= \bra{\pm X} \hat{P} \ket{\mp X}
		,\end{align*}
		while the second solution is only valid for
		\begin{align*}
			\bra{\pm\frac{\pi}{2}, 0} \hat{P} \ket{\mp\frac{\pi}{2}, 0} 
			&= \bra{\pm\frac{\pi}{2}, \pi} \hat{P} \ket{\mp\frac{\pi}{2}, \pi} \\
			&= \bra{\pm X} \hat{P} \ket{\mp X}
		. \end{align*}
	
		The solutions can be explicitly expressed as
		\[
			\bra{\pm X} \hat{P} \ket{\mp X}
			= (\pm i)^{N_Z} \delta_{N_X, 0}\delta_{N_I, 0}
		,\]
		where we assumed $N$ is even.

	\item {$|\beta_{i,j}|=|\kappa_{i,j}|=1$, $\alpha_{i,j} = \gamma_{i,j} = 0$.}
		
		Solving \cref{eq:a1,eq:b1,eq:c1,eq:d1} under these constrains yields the solution
		$a = 1-b, c = \frac{b-1}{2b - 1}, d = \frac{b}{2b-1}$.  
		By the domain constraint $b \in [0,1]$ it can be found that
		$(c,d) \in \{(1,0), (0,1)\}$, where the first result happens for $b=0$ and the second for $b=1$.
		Thus, we obtain two proper solutions: $a = 1, b = 0, c = 1, d = 0$ and $a = 0, b = 1, c = 0, d = 1$.
		From previous results we know 
		the first solution is only valid for 
		\begin{align*}
			\bra{\frac{\pi}{2}, \pm\frac{\pi}{2}} \hat{P} \ket{-\frac{\pi}{2}, \mp \frac{\pi}{2}} 
			&= \bra{-\frac{\pi}{2}, \pm\frac{\pi}{2}} \hat{P} \ket{\frac{\pi}{2}, \mp \frac{\pi}{2}} \\
			&= \bra{\pm Y} \hat{P} \ket{\pm Y}
		,\end{align*}
		while the second solution is only valid for
		\begin{align*}
			\bra{\frac{\pi}{2}, \pm\frac{\pi}{2}} \hat{P} \ket{\frac{\pi}{2}, \pm \frac{\pi}{2}} 
			&= \bra{-\frac{\pi}{2}, \pm\frac{\pi}{2}} \hat{P} \ket{-\frac{\pi}{2}, \pm \frac{\pi}{2}} \\
			&= \bra{\pm Y} \hat{P} \ket{\pm Y}
		.\end{align*}
		
		The solutions can be explicitly expressed as
		\[
			\bra{\pm Y} \hat{P} \ket{\pm Y}
			= (\pm 1)^{N_Y} \delta_{N_X, 0}\delta_{N_Z, 0}
		.\]

	\item {$|\gamma_{i,j}|=|\kappa_{i,j}|=1$, $\alpha_{i,j} = \beta_{i,j} = 0$.}
		
		Solving \cref{eq:a1,eq:b1,eq:c1,eq:d1} under these constrains yields two solutions.

		The first solution, $c = d = 1$, 
		implies $\theta_i=\theta_j \in \{\pi,0\}$.
		Again, we impose $\phi_i = \phi_j = 0$ for the poles.
		This solution is only valid for 
		\begin{align*}
			\bra{0, 0} \hat{P} \ket{0, 0} 
			&= \bra{Z} \hat{P} \ket{Z}, \\
			\bra{\pi, 0} \hat{P} \ket{\pi, 0} 
			&= \bra{-Z} \hat{P} \ket{-Z}.
		\end{align*}
		
		The second solution, $a = b = \frac{1}{2}, c = 2-d$, 
		is identical to the previous solution with unnecessary constraints on $a, b$; since $c, d\in[0,1]$ implies $d = c = 1$.

		The solutions can be explicitly expressed as
		\[
			\bra{\pm Z} \hat{P} \ket{\pm Z}
			= (\pm 1)^{N_Z} \delta_{N_X, 0}\delta_{N_Y, 0}
		.\]

\end{itemize}

We can summarize the obtained results as:
\begin{equation}\label{eq:pauli_deltas}
\begin{cases}
	\bra{\pm Z} \hat{P} \ket{\mp Z} &= (\pm i)^{N_Y}\delta_{N_Z, 0}\delta_{N_I, 0}, \\
	\bra{\pm Y} \hat{P} \ket{\mp Y} &= (\pm i)^{N_Z}\delta_{N_Y, 0}\delta_{N_I, 0}, \\
	\bra{\pm X} \hat{P} \ket{\pm X} &= (\pm 1)^{N_X}\delta_{N_Y, 0}\delta_{N_Z, 0}, \\
	\bra{\pm X} \hat{P} \ket{\mp X} &= (\pm i)^{N_Z}\delta_{N_X, 0}\delta_{N_I, 0}, \\
	\bra{\pm Y} \hat{P} \ket{\pm Y} &= (\pm 1)^{N_Y}\delta_{N_X, 0}\delta_{N_Z, 0}, \\
	\bra{\pm Z} \hat{P} \ket{\pm Z} &= (\pm 1)^{N_Z}\delta_{N_X, 0}\delta_{N_Y, 0}, 
\end{cases}
\end{equation}
where every other matrix element is dependent on the system size $N$.

In conclusion, we obtain that only spin coherent states along the $\pm X, \pm Y, \pm Z$ directions in the Bloch spheres contribute to the Pauli spectra without influence from the system size $N$ and that such contributions only arise from terms in \cref{eq:pauli_arbitrary_sup} that involve either a single spin coherent state along one of this directions or two of such states that are orthogonal to each other, i.e. two antipodal points such as $+Z$ and $-Z$ directions.
Moreover, these relevant states are also the only stabilizer states available in the permutation invariant sector, which hints at a possible generalization to stabilizer states in larger Hilbert spaces that include lower total spin states, or other symmetries.

\section{Approximated Stabilizer Rényi Entropy in the permutation invariant sector}
\label{sup:large_N}

Based on the results \labelcref{eq:pauli_deltas}, we propose 
to reduce the continuous integration domain in \cref{eq:pauli_arbitrary_sup} to compute the Pauli spectra in terms of a limited number of matrix elements.
We now provide a detailed derivation of our main result, an approximation of the SRE which is valid for large $N$ in the permutation invariant sector, \cref{eq:SRE_approx_sup}. In practice, 
since
\begin{equation*}
	\frac{2J+1}{4\pi}\int d\vb*{\Omega}' \ket{\theta',\phi'}\bra{\theta',\phi'}\ket{\theta,\phi}
	= \hat{\mathbb{I}}_{J} \ket{\theta, \phi} 
	= \ket{\theta, \phi},
\end{equation*}
we turn the double integral in \cref{eq:pauli_arbitrary_sup} into a sum over the appropriate matrix elements described in \labelcref{eq:pauli_deltas} while removing the normalization factor $[(2J+1)/(4\pi)]^2$.
Thus, we obtain
\begin{equation*}
	\lim_{N \to \infty} \ev*{\hat{P}} 
	\simeq 
	\ev*{\hat{P}}_\mathrm{STAB}
,\end{equation*}
with
\begin{equation}
\begin{split}
	\ev*{\hat{P}}_\mathrm{STAB}
	\equiv \sum_{\sigma} \bigg[ 
    &(-1)^{N_{\overline{\sigma}}} \bra{\psi}\ket{-\sigma}\bra{\sigma}\ket{\psi} \bra{\sigma}\hat{P}\ket{-\sigma}\\
	  &+\bra{\psi}\ket{\sigma}\bra{-\sigma}\ket{\psi} \bra{\sigma}\hat{P}\ket{-\sigma} \\
	  &+ |\bra{\sigma}\ket{\psi}|^2\bra{\sigma}\hat{P}\ket{\sigma} \\ 
    &+ (-1)^{N_\sigma}|\bra{-\sigma}\ket{\psi}|^2 \bra{\sigma}\hat{P}\ket{\sigma}
 \bigg],
\end{split}
\end{equation}
where $\sigma\in\{X,Y,Z\}$,and $\overline{\sigma}$ corresponds to any other $\overline{\sigma} \ne \sigma$. More explicitly, 
\begin{equation}
	\begin{split}
		\ev*{\hat{P}}_\mathrm{STAB} = 
        & \quad c_{X,1}\delta_{N_Y, 0}\delta_{N_Z, 0} 
		+ c_{X,2}\delta_{N_X, 0}\delta_{N_I, 0} \\
		&+ c_{Y,1}\delta_{N_Y, 0}\delta_{N_Z, 0} 
		+ c_{Y,2}\delta_{N_Y, 0}\delta_{N_I, 0} \\
		&+ c_{Z,1}\delta_{N_X, 0}\delta_{N_Z, 0} 
		+ c_{Z,2}\delta_{N_Z, 0}\delta_{N_I, 0}
	.\end{split}
\end{equation}
where 
\begin{align}
	c_{X,1} =& \sum_{k=0}^{\infty} \left(c^{(X)}_{1,2}\delta_{N_X, 2k} + c^{(X)}_{1,1}\delta_{N_X,2k+1} \right)  \label{eq:c_x_1}\\
	c_{X,2} =&\sum_{k=0}^{\infty} (-1)^{k} \left( c^{(X)}_{2,2}\delta_{N_Z, 2k} + c^{(X)}_{2,1}\delta_{N_Z,2k+1}\right) \\
	c_{Y,1} =&\sum_{k=0}^{\infty} \left( c^{(Y)}_{1,2}\delta_{N_Y, 2k} + c^{(Y)}_{1,1}\delta_{N_Y,2k+1}\right) \\
	c_{Y,2} =&\sum_{k=0}^{\infty} (-1)^{k} \left( c^{(Y)}_{2,2}\delta_{N_Z, 2k} + c^{(Y)}_{2,1}\delta_{N_Z,2k+1}\right) \\
	c_{Z,1} =&\sum_{k=0}^{\infty} \left( c^{(Z)}_{1,2}\delta_{N_Z, 2k} + c^{(Z)}_{1,1}\delta_{N_Z,2k+1}\right) \\
	c_{Z,2}	=&\sum_{k=0}^{\infty} (-1)^{k} \left( c^{(Z)}_{2,2}\delta_{N_Y, 2k} + c^{(Z)}_{2,1}\delta_{N_Y,2k+1}\right)  \label{eq:c_z_2}
,\end{align}
with
\begin{align}
	c^{(\sigma)}_{1,m} =& |\bra{\sigma}\ket{\psi}|^2 +(-1)^m |\bra{-\sigma}\ket{\psi}|^2, \\
    \begin{split}
	c^{(\sigma)}_{2,m} =& \left( \bra{\psi}\ket{-\sigma}\bra{\sigma}\ket{\psi} + (-1)^m \bra{\psi}\ket{\sigma}\bra{-\sigma}\ket{\psi} \right) \\
    &\cdot\sqrt{(-1)^m}
    .\end{split}
\end{align}
As the expectation value is then given by these six terms, we can expand $\ev*{\hat{P}}^{2q}$ using the multinomial theorem as
\begin{equation*}
	\begin{split}
	\ev*{\hat{P}}^{2q}_\mathrm{STAB} 
    = \sum_{\sum k_{i} = 2q} &
	{2q \choose {k_1, k_2, k_3, k_4, k_5, k_6}} \\
	&\cdot c_{X,1}^{k_1} c_{X,2}^{k_2} c_{Y,1}^{k_3} c_{Y,2}^{k_4} c_{Z,1}^{k_5} c_{Z,2}^{k_6} \\
    &\cdot \delta_{N_X, 0}^{k_2+k_3+k_5}\delta_{N_Y, 0}^{k_1+k_4+k_5}\\
    &\cdot \delta_{N_Z, 0}^{k_1+k_3+k_6}\delta_{N_I, 0}^{k_2+k_4+k_6}
	.\end{split}
\end{equation*}

We now select the terms that can be different than zero based on the Pauli string constraint $N_X+N_Y+N_Z+N_I = N$ and $\delta_{n,0}^0 = 1$.

\begin{equation}
	\label{eq:pauli_spectra_2q}
	\begin{split}
		\ev*{\hat{P}}^{2q}_\mathrm{STAB}
		= 
        &\quad c_{X,1}^{2q}\delta_{N_Y, 0}\delta_{N_Z, 0} + c_{X,2}^{2q}\delta_{N_X, 0}\delta_{N_I, 0}\\
		  &+ c_{Y,1}^{2q}\delta_{N_X, 0}\delta_{N_Z, 0} + c_{Y,2}^{2q}\delta_{N_Y, 0}\delta_{N_I, 0}\\
		  &+ c_{Z,1}^{2q}\delta_{N_X, 0}\delta_{N_Y, 0} + c_{Z,2}^{2q}\delta_{N_Z, 0}\delta_{N_I, 0}\\
		  &+ \delta_{N_X, N} (c_{X,1} + c_{Y,2} + c_{Z,2})^{2q}  \\
        &- \delta_{N_X, N} \left( c_{X,1}^{2q} + c_{Y,2}^{2q} + c_{Z,2}^{2q} \right) \\
		  &+ \delta_{N_Y, N} (c_{X,2} + c_{Y,1} + c_{Z,2})^{2q} \\
        &- \delta_{N_Y, N} \left( c_{X,2}^{2q} + c_{Y,1}^{2q} + c_{Z,2}^{2q} \right) \\
		  &+ \delta_{N_Z, N} (c_{X,2} + c_{Y,2} + c_{Z,1})^{2q} \\
		  &- \delta_{N_Z, N} \left( c_{X,2}^{2q} + c_{Y,2}^{2q} + c_{Z,1}^{2q} \right) \\
		  &+ \delta_{N_I, N} (c_{X,1} + c_{Y,2} + c_{Z,1})^{2q} \\
        &- \delta_{N_I, N} \left( c_{X,1}^{2q} + c_{Y,2}^{2q} + c_{Z,1}^{2q} \right) 
	,\end{split}
\end{equation}
where we have used the relation
\begin{equation*}
	 \sum_{\substack{\sum k_i=n \\ k_1+k_2 > 0 \\ k_2+k_3>0 \\ k_1+k_3>0}} {n \choose {k_1,k_2,k_3} } a^{k_1} b^{k_2} c^{k_3}
	 = (a+b+c)^n - a^n - b^n - c^n
.\end{equation*}

Each of the coefficients $c_{\sigma, m}, \forall \sigma \in \{X,Y,Z\}, m\in\{1,2\}$ reduce to a single term depending on the parity of the Pauli string of interest, so we can sum over the Pauli spectra in \cref{eq:pauli_spectra_2q} by separating each case in even and odd terms following \crefrange{eq:c_x_1}{eq:c_z_2}. For example,
\begin{equation*}
	\begin{split}
	\sum_{\hat{P}} c_{X,1}^{2q}\delta_{N_Y, 0}\delta_{N_Z, 0} 
	=& \sum_{N_X=0}^{N} {N \choose N_X} c_{X,1}^{2q} \\ 
	=& \left[\sum_{2k=0}^{N} {N \choose 2k}  \right]  \left(c^{(X)}_{1,2}\right)^{2q} \\
	&+ \left[\sum_{2k+1=1}^{N-1} {N \choose 2k+1} \right] \left(c^{(X)}_{1,1}\right)^{2q} \\ 
	=& 2^{N-1}\left[ \left(c^{(X)}_{1,2}\right)^{2q} + \left(c^{(X)}_{1,1}\right)^{2q}\right]
	.\end{split}
\end{equation*}

The last four terms in \cref{eq:pauli_spectra_2q} will be non-zero only for a particular Pauli string each, 
and we may safely ignore them with respect to the previous terms, as $|\ev*{\hat{P}}| \le 1;\forall \hat{P}$ and $D = 2^N \gg 1$. With this, we find the approximate SRE of a state in the maximal total spin manifold for a sufficiently large $N$ to be
\begin{equation}\label{eq:SRE_approx_sup}
	\begin{split}
	\lim_{N \to \infty} \mathcal{M}_q 
	&= \lim_{N \to \infty} \frac{1}{1-q}\log_2\left( \frac{\sum_{\hat{P}} \ev*{\hat{P}}^{2q}}{D} \right) \\
	&\simeq \frac{1}{1-q}\log_2\left( \frac{1}{2}
    \sum_{\sigma, n, m}
    \left(c_{n,m}^{(\sigma)}\right)^{2q} \right),
    \end{split}
\end{equation}
where ${\sigma\in\{X,Y,Z\}}$, ${n\in\{1,2\}}$, ${m\in\{1,2\}}$.

\section{Many-body Bell inequality}\label{app.bell}
To quantify the strength  of many-body Bell correlations generated in the OAT process, we use a broad family of Bell inequalities first introduced in 
Refs~\cite{CavalcantiPRL2007,CavalcantiPRA2011,HePRA2011,Chwedeczuk2022,Plodzien2024PRA}. When each of $N$ parites measures two binary quantities $\sigma_{1,2}^{(k)}=\pm1$ (with $k=1\ldots N$),
then the correlator
\begin{align}
 \tilde{ \mathcal E}=\modsq{\av{\sigma_+^{(1)}\ldots\sigma_+^{(N)}}},
\end{align}
with $\sigma_+^{(k)}=\frac12(\sigma_1^{(k)}+i\sigma_2^{(k)})$ can be reproduced by a theory consistent with the postulates of local realism if it takes the form
\begin{align}
  \tilde{\mathcal E}=\modsq{\int\!\! d\lambda\, p(\lambda)\sigma_+^{(1)}(\lambda)\ldots\sigma_+^{(N)}(\lambda)},
\end{align}
where $\lambda$ is a hidden variable and $p(\lambda)$ is its probability distribution. Using the Cauchy-Schwarz inequality we obtain
\begin{align}\label{eq.bell.ineq}
  \tilde{\mathcal E}\leqslant\int\!\! d\lambda\, p(\lambda)\modsq{\sigma_+^{(1)}(\lambda)\ldots\sigma_+^{(N)}(\lambda)}=\frac1{2^N},
\end{align}
which is the $N$-body Bell inequality. 

For quantum systems, $\sigma_{1,2}^{(k)}$ are replaced with the Pauli operators $\hat\sigma_{1,2}^{(k)}$ and $\tilde{\mathcal E}$ is replaced by $\mathcal E$, which is a quantum correlator, i.e., 
\begin{align}
  \mathcal E =\modsq{\av{\hat\sigma_+^{(1)}\ldots\hat\sigma_+^{(N)}}},
\end{align}
where $\hat\sigma_+^{(k)}$ is a rising operator. If $\mathcal E$ violates the bound from Eq.~\eqref{eq.bell.ineq}, it witnesses the many-body Bell correlations. 

For bosonic systems, qubits cannot be addressed individually, thus the $\hat\sigma_+^{(k)}$'s must be replaced with 
collective angular momentum operators. Formally, this is achieved by symmetrizing the product of $N$ operators $\hat\sigma_+^{(1)}\ldots\hat\sigma_+^{(N)}$. Since
all orderings of these operators are equivalent and there are $N!$ such settings, the Bell inequality~\eqref{eq.bell.ineq} for bosonic systems takes the form
\begin{equation}    
\begin{split}
  \mathcal E&=\modsq{\frac1{N!}\av{\hat J_+^N}}\leqslant 2^{-N},\\
  Q & = \log_2(2^N {\mathcal E}) \leqslant 0,
  \end{split}
\end{equation}

The many body correlator ${\mathcal E}$ captures the structure of the many-body quantum state.
For instance, when the state is $2-$local, the maximal value of the correlator is reached when $N-2$ parties form a quantum-mechanical GHZ state, giving the value of the correlator ${\mathcal E}=\frac14$, $Q = N-2$,
while the remaining one is
a particle measurements on which yield binary outcomes. In this case, the correlator is 
${\mathcal E}=\frac14\cdot\frac12$.
Naturally, a $2-$local state can have weaker correlations. For example, when the state is the product of two GHZ states of $N/2$ particles each, then ${\mathcal E}=\frac14\cdot\frac14$.
When the system is $3-$local, then again the correlator is maximal in the scenario, when $N-2$ particles form a GHZ state and the remaining two are mutually Bell-uncorrelated. 
 As an another example we can consider a $N$-spins state being a product of $n$ separable parties, each producing binary outcomes and not restricted by quantum mechanics, and $k$ products of GHZ states, then ${\mathcal E}=\frac{1}{2^n}\cdot\frac{1}{4^k} = \frac{1}{2^{2k+n}}$. 
 Bounds for $Q$ can be generalized to states being a
product of $n$ single qubits and $(N-n)$-entangled state $
  N-1-(n+1) < Q \le N-1-n$, and the correlator $Q$ indicates the existence of entanglement depth
$d_e = N-n$. In general, for $k$-separable states the correlator $Q$ is
bounded from above
$Q \le N - k$,
and violation of this inequality witnesses a non-$k$-separability,
i.e. the state is a product of at most $(k-1)$-entangled states, for details see \cite{Plodzien2024PRA}.

\bibliography{biblio} 

\end{document}